%% file: main.tex
\documentclass[a4paper]{report}
\usepackage[english]{babel}
\usepackage[utf8x]{inputenc}
\usepackage[T1]{fontenc}
\usepackage{pdfpages}
\usepackage{amssymb}
\usepackage{amsthm}
\usepackage{xcolor}
\usepackage[a4paper,top=3cm,bottom=2cm,left=3cm,right=2.5cm,marginparwidth=1.75cm]{geometry}
\usepackage{amsmath}
\usepackage{graphicx}
\usepackage{algorithm}
\usepackage[noend]{algpseudocode} 
\usepackage[colorinlistoftodos]{todonotes}
\usepackage[colorlinks=true, allcolors=blue]{hyperref}
\usepackage{subcaption}
\usepackage[onehalfspacing]{setspace}
\usepackage{booktabs}

\newtheorem*{theorem}{Theorem}
\newtheorem*{Lemma}{Lemma}
\newtheorem*{definition}{Definition}
\newtheorem*{conjecture}{Conjecture}
\newtheorem*{question}{Question}

\begin{document}
\input{title.tex}

\begin{abstract}
We are living in a world which is getting more and more interconnected and, as physiological effect, the interaction between the entities produces more and more information. This high throughput generation calls for techniques able to reduce the volume of the data, but still able to preserve the carried knowledge. Data compression and summarization techniques are one of the possible approaches to face such problems. The aim of this thesis is to devise a new pipeline for compressing and decompressing a graph by exploiting Szemerédi’s Regularity Lemma. In particular, it has been developed a procedure called CoDec (Compression-Decompression) which is based on Alon et al's constructive version of the Regularity Lemma. We provide an extensive experimental evaluation to measure how robust is the framework as we both corrupt the structures carried by the graph and add noisy edges among them. The experimental results make us confident that our method can be effectively used as a graph compression technique able to preserve meaningful patterns of the original graph.



\end{abstract}


\begin{center}
\subsection*{Acknowledgements}
I would like to express my gratitude to my supervisor Marcello Pelillo for accepting me as a candidate and assigning me such stimulating research topic for my thesis.\\[5mm]

I also have to express my gratitude to Marco Fiorucci as he played a crucial role between me and my supervisor. He guided me during the creation of this thesis and he was always kind enough to discuss all the problems faced and to improve all the ideas that I had.\\[5mm]

Since this thesis ends an important period of my life, I have to thank both my old and new friends that I've met, for all the happy moments and beautiful discussions that we had.\\[5mm]

I have to thank my family for the continuous moral and financial support during these two years. In particular, my brother and his family to always have kind words for me in both happy and difficult times. Last but not least, I have to thank my parents who always constituted a source of love for me and that allowed me to pursue this degree.
\end{center}

\tableofcontents

\newpage

\chapter{Introduction}\label{chap:introduction}
\input{introduction.tex}

\chapter{Context and Background}\label{chap:background}
\input{background.tex}

\chapter{Description of the devised technique}\label{chap:devisedtechnique}
\input{devisedtechnique.tex}

\chapter{Experimental Results}\label{chap:experimentalresults}
\input{experimentalresults.tex}

\chapter{Conclusions}\label{chap:conclusion}
\input{conclusion.tex}

\bibliographystyle{ieeetr}
\bibliography{references.bib,dsets.bib}

\end{document}

%% file: title.tex
\begin{titlepage}
\begin{center}
	\LARGE{Ca' Foscari University of Venice}\\
	\vspace*{3mm}
	\large
    \textsc{\Large Department of Environmental\\Sciences, Informatics and Statistics}\\

	\vspace*{5mm}
	\begin{figure}[ht] 
	\center \includegraphics[height=50mm]{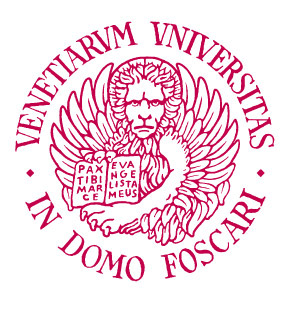}
	\end{figure}
	\vspace*{5mm}
	Master's Degree programme in\\ 
	Computer Science, Data Management and Analytics\\
	Second Cycle (D.M. 270/2004)\\
	\rule{\linewidth}{1pt}
	\vspace*{10mm}
	M.Sc. Thesis\\
	\vspace*{10mm}
	\huge{Graph Compression Using The Regularity Method}
\end{center}
\vspace*{20mm}
\begin{flushleft}
	\textsc{Supervisor:}\hspace{\stretch{2}}\textsc{Graduand:}\\
	Prof. Marcello Pelillo \hspace{\stretch{1}}Francesco Pelosin\\\hspace{\stretch{1}}Matr. 839220\\
	\textsc{Assistant supervisor:}\\
	Dr. Marco Fiorucci\\
	\vspace*{3mm}
\end{flushleft}
\vspace*{25mm}
\normalsize
\begin{center}
	Academic Year \\2016 - 2017
\end{center}

\vfill
\end{titlepage}

%% file: introduction.tex
We are living in a world which is getting more and more interconnected and, as physiological effect, the interaction between the entities produces more and more information. A structure to store, represent and manage such information is a graph. Graphs are mathematical objects consisting in a vertex set $V$ which represents the entities and an edge set $E$ representing the links between them. Graphs are ubiquitous, representing a variety of natural processes as diverse as friendships between people, communication patterns and interactions between neurons in the brain. Often the chosen data structure to represent a graph is its adjacency matrix $A$, which can be binary, corresponding to whether there exists an interaction between two vertices, or numerical, corresponding to the strength of the connection. 

This high throughput generation calls for techniques able to reduce the volume of the data but still able to preserve the knowledge that carry. The data mining community has taken a strong interest in the task since devising techniques that are able to cope with the intrinsic complexity and dimension of data imply several advantages. The trivial advantage is the reduction of data volume and storage but we can also store only meaningful patterns by filtering the noise which data naturally have. Data compression and summarization techniques are one of the possible approaches to face such problems. The survey of Liu et al \cite{Liu2017GraphSurvey} provides a comprehensive overview of the state-of-the-art methods for summarizing graph data. The essence of such techniques is to transform the input graph to a more concise representation by preserving the essential patterns which carry the most relevant information to describe the phenomenon under study.

In particular, we exploit the power of Szemerédi's Regularity Lemma to devise a well-found framework for compressing large graphs. The Regularity Lemma was introduced by the Abel laureate Endre Szemerédi in the mid 70's as a tool to prove his famous theorem on arithmetic progressions. This tool roughly states that we can approximate every large graph by partitioning it into a small number of parts such that the bipartite subgraph between almost every pair of parts is random-like. The Lemma guarantees that the partitioning, which in fact represent the compression of the graph, when properly decompressed, preserves some topological properties of the input graph. 

The first formulation of the Regularity Lemma is an existence predicate, Szemerédi proved the existence of a regular partition but he did not provide a method to obtain it. In the last decades several mathematicians devised constructive versions of the Regularity Lemma. However, all these algorithms hide a tower constant that makes the implementation unfeasible.

Sperotto and Pelillo \cite{SperottoPelillo} reported arguably the first practical application of the Regularity Lemma, the original motivation was to study how to take advantage of the partitions in a pairwise clustering context. In particular, they proposed the first approximate version of the exact algorithm introduced by Alon et al. \cite{Alon94}. 

In the past few years different approximate algorithms, explicitly inspired by the Regularity Lemma, have been applied in pattern recognition, bioinformatics and social network analysis. The reader can refer to \cite{PelilloRevealing17} for a survey of these emerging techniques. The last improved version based on the Alon et al. algorithm has been introduced by Fiorucci et al. \cite{Fioruccietal} where they improved the work proposed by Sperotto and Pelillo by introducing new heuristics. This thesis tries to further improve and investigate the conditions of application of the proposed algorithm.

\section{Contributions}\label{sec:aim}

This thesis aims to improve and further investigate the conditions of application of Fiorucci et al.'s approximate version \cite{Fioruccietal} of Alon et al.'s algorithm \cite{Alon94}. In particular, the first contribution of this thesis relies on the introduction of a new heuristic for the refinement step which exploits the internal properties of the classes composing a regular partition. The heuristic idea relies on intuitive motivation which has its roots on the combinatorial nature of the lemma. It tries to mitigate the required tower bound proved by Gowers \cite{Gowers1997} on the number of the classes composing a regular partition, making, as we will see, the algorithmic implementation able to find partitions even in medium graphs.

The second and even more important contribution is the definition of a pipeline that we called CoDec. The pipeline first compresses a graph using our new version of Alon et al.'s algorithm, and then decompresses the obtained structure into another graph. After the decompression it exploits a new post-decompression step which aims to improve the structural preservation quality. 

The work then points out the limits and possible strengths of the technique through an extensive batch of experiments both conducted over synthetic graphs and real networks. We will then analyze the results and draw some considerations on the devised technique.

Last but not least, the outline of the thesis provides, as much as it could, a progressive and concise dive into such deep topic. Several figures and simple conceptual explanations of the mathematical formulas which constitutes the nature of this powerful lemma, have been reported to offer an accessible reading to whom approaches this topic for the first time. The work deliberately does not report all the proofs of the theorems and lemmas cited since we can find them in the original papers, a reference will point us to the correct resource.


\section{Structure of the work}\label{sec:structure}
This thesis has been divided in chapters each of which tries to give a full description of the topic discussed by giving the essential informations needed to understand the core idea. In particular:

\begin{itemize}

\item In Chapter \ref{chap:background} we will briefly see the two main mathematical areas where Szemerédi's Regularity Lemma originated. It is important to understand the mathematical nature of the lemma to fully grasp its details. Then, we will progressively dive into its specifications until we will be able to fully state it and complete it with the correct details. After understanding the origins and the power of Szemerédi's Regularity Lemma, we will study Alon et al.'s algorithm which basically is the reformulation of the lemma in algorithmic terms, that is the algorithm that we exploited and tweaked to constitute the compression step of the CoDec procedure. 

\item In Chapter \ref{chap:devisedtechnique} we will describe the CoDec procedure in all its steps and specifications. In particular, we will give a brief overview of the main steps accompanied with several sketch figures to clarify the logic behind. Then we will describe all  its steps that are Compression (we will only focus on the new heuristic developed since the algorithm is explained in Section \ref{sec:alon}), Decompression and Post-Decompression Filtering.

\item In Chapter \ref{chap:experimentalresults} we will discuss the experimental results. Firstly we define the measures used to validate the experiments, then we will move on to describe how the synthetic data is created and where real data is taken from in order to test the framework. We will provide an exhaustive description of the experiments and we will report them through several images and tables. Then we perform some analysis and make some considerations of the results obtained. 

\item In Chapter \ref{chap:conclusion} we will draw some conclusions that emerged from the experimental results. Then we will point out further investigations that could improve the framework.

\end{itemize}

%% file: background.tex
In this section we will see the background needed to understand the nature of the technique devised during this work, in particular we will briefly describe and illustrate some examples of the problem treated in the mathematical areas from which Szemerédi's Regularity Lemma comes from. It is very important to point out the the origins of the method exploited in our work since it gives the right philosophy in order to think in the correct terms. 

\section{Extremal Graph Theory and Ramsey Theory}\label{sec:extremal}

Szemerédi's Regularity Lemma is one of the most powerful tools in Extremal Graph Theory and it is widely used by mathematicians to prove theorems which require to demonstrate the existence of some substructure in a graph. In fact these kind of questions are the questions mainly posed by these two mathematical areas: Extremal Graph Theory and Ramsey Theory. 

\subsection*{Extremal Graph Theory} Extremal Graph Theory is a sub-area of Combinatorics which focuses on the study of any discrete or finite system, in particular, the mathematical objects treated are finite Sets, Colourings and  Graphs. It is maybe one of the oldest subareas of Combinatorics and an example of the problems that treats is:

\begin{question}
What is the maximum number of edges in a triangle-free graph of $n$ vertices?
\end{question}

\begin{figure}
\centering
\includegraphics[width=0.5\textwidth]{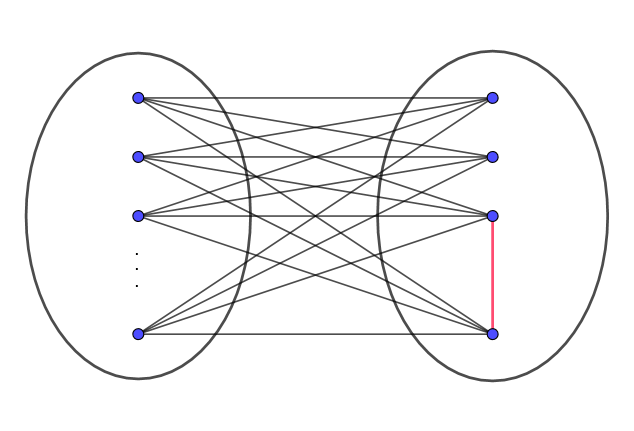}
\caption{\label{fig:bipartite}A complete bipartite graph, it can be seen that if we add an edge (red) we are creating a triangle.}
\end{figure}

A complete bipartite graph (with part sizes as equal as possible) has $\left \lfloor \frac{n}{2} \right \rfloor \left \lceil \frac{n}{2} \right \rceil \approx \frac{n^2}{4}$ edges and no triangle, this result comes from Mantel's theorem which has been proven in 1907 and later generalized by Turán in 1943. The proof is pretty simple and a visual representation is given by Figure \ref{fig:bipartite}. If we add a single edge in a complete bipartite graph we are, in fact, creating a triangle, so the maximum number of edges of a triangle free graph is the one given by Mantel. More generally the objective of Extremal Graph Theory is to suppose the existence (or not) of a certain sub-structure and then discover how large the graph should be for that sub-structure to exists (or not). 

\subsection*{Ramsey Theory} A closely related field of Extremal Graph theory is Ramsey Theory. What it basically aims to do is finding ``order in chaos''. It is slightly more general than Extremal Graph Theory and the questions which asks treat, not only graphs, but also different mathematical objects.

We could roughly describe Ramsey Theory as the collection of problems where you have some Coloring of some object and you're looking for some very structured subset. If a given object is very very large and colored in an arbitrary way however it is colored you will find the the sub-object you're looking for. Since the Regularity Lemma has been used as a tool to prove the main Szemerédi Theorem which treats arithmetic progressions, we will illustrate a simple example of the problems studied in Ramsey Theory using arithmetic progressions, but before seeing it let's give two definitions: 

\begin{definition}
An arithmetic progression is a sequence of $l$ natural numbers of constant difference $d$ with an initial value $i$.
\end{definition}

For example the progression with $l=5, d=7, i=23$ is $23,30,37,44,51$.

\begin{definition}
A colouring of a non-empty set $S$ with $r \in \mathbb{N}$ colours is just a colourful way of denoting a map $c : S \rightarrow [r]$. The number of $\{1,\dots,r\}$ are usually the colours, and $s \in S$ is painted with colour $i\in [r]$ if $c(s)=i$
\end{definition}

\paragraph{Example}Now we are ready to illustrate the example: By colouring the natural numbers using only two colours, say \textcolor{red}{red} and \textcolor{blue}{blue}, it is easy to see that 9 consequent numbers e.g. $1,2,\dots,9$, are needed to ensure that there is an arithmetic progression of length 3. Why? Let us try to prove the opposite, so suppose the sequence $1,2,\dots,9$ does not contain arithmetic progressions of length 3. For this reason, number 1, 5 and 9 cannot be all equally coloured. So assume first that \textcolor{red}{1}  and \textcolor{red}{5} are red, and \textcolor{blue}{9} is blue. Since \textcolor{red}{1} and \textcolor{red}{5} are red, \textcolor{blue}{3} has to be blue. But \textcolor{blue}{9} is also blue, so \textcolor{red}{6} has to be red. Now \textcolor{red}{5} and \textcolor{red}{6} are red, forcing \textcolor{blue}{4} and \textcolor{blue}{7} to be blue. Number \textcolor{red}{8} has to be red since \textcolor{blue}{7} and \textcolor{blue}{9} are blue, and number \textcolor{red}{2} has to be red since \textcolor{blue}{3} and \textcolor{blue}{4} are blue. But then \textcolor{red}{2}, \textcolor{red}{5} and \textcolor{red}{8} are all red, which is a contradiction. The case where 1 and 9 are red and 5 is blue is treated similarly. On the other hand, the sequence of length 8, given by \textcolor{red}{R}\textcolor{blue}{B}\textcolor{red}{R}\textcolor{blue}{B}-\textcolor{blue}{B}\textcolor{red}{R}\textcolor{blue}{B}\textcolor{red}{R} has no arithmetic progressions of length 3. Thus 9 is a sharp bound for this property. This is also called van der Waerden number W(2,3)=9.

\section{Szemerédi's Regularity Lemma}\label{sec:szemeredilemma}

Szemerédi's Regularity Lemma arises from the composition of the two above areas, as we said it is a powerful lemma which Szermerédi used as a tool to prove his famous theorem on the arithmetic progressions, which let him win the prestigious Abel Prize in 2012. The theorem is a generalization of a much older problem posed by Erd\"{o}s and Tur\'{a}n in 1936.

By defining upper density of a set $A \subset \mathbb{N}$ as:

$$\tilde{d}(A) = \underset{n \to \infty}{limsup}\frac{|A \cap [n]|}{n}$$

and by denoting $[n] = \{1, \dots, n\}$ the set of all positive integers from 1 to $n$, the conjecture says:

\begin{conjecture}[Erd\"{o}s and Tur\'{a}n, 1936]
If $\tilde{d}(A) > 0$, then $A$ contains arbitrarily long arithmetic progressions.
\end{conjecture}

In 1953, Klaus Friedrich Roth proved that any subset of the integers with positive upper density contains an arithmetic progression of length 3. In 1969, Endre Szemerédi proved that the subset must contain an arithmetic progression of length 4, and then in 1975 proved that any subset with positive upper density must contain arithmetic progressions of arbitrary length, known as Szemerédi's Theorem \cite{Szemeredi1969OnProgression}. The theorem has been proven in several different ways for example Furstenberg \cite{Furstenberg1982TheTheorem} proved it through Ergodic Theory, Gowers \cite{Gowers1998} using Harmonic Analysis and many other, according to Terry Tao there are at least 16 different ways to prove it. Each of these proves really opened a new area of mathematics, so this conjecture is sometimes described as: \textit{``the greatest triumph of Hungarian school of mathematics''}, which says that if you ask the right question, and try to solve it, the connections between different areas will show themselves and you will discover the theory anyway instead of building it. 

In order to prove his famous theorem, which is very intricate, Szemerédi had to introduce several other mathematical lemmas and tools, the Regularity Lemma is one of these. The Lemma can be put into words with the following statement: 

\paragraph{Vague Version}\textit{Every graph $G$ can be ``well approximated'' by a bounded number of ``quasi-random graphs''}\\

Surely any graph does not look like a random graph, but maybe any graph looks like a combination of random graphs, that's what he claimed and proved. The interesting thing is that both the approximation of the graph and the amount of randomness of the random graphs only depend by one parameter $\epsilon$. The number of graphs needed is $k(\epsilon)$ a function of $\epsilon$. In order to understand it better we introduce another less vague version.

\paragraph{Less Vague Version}\textit{Given any graph $G$, we can partition the vertex set $V(G)$ into a bounded number of classes (at most $k$, say), such that ``almost every'' pair $(A, B)$ is well approximated by a quasi-random bipartite graph.
}\\

\begin{figure}
\centering
\includegraphics[width=0.7\textwidth]{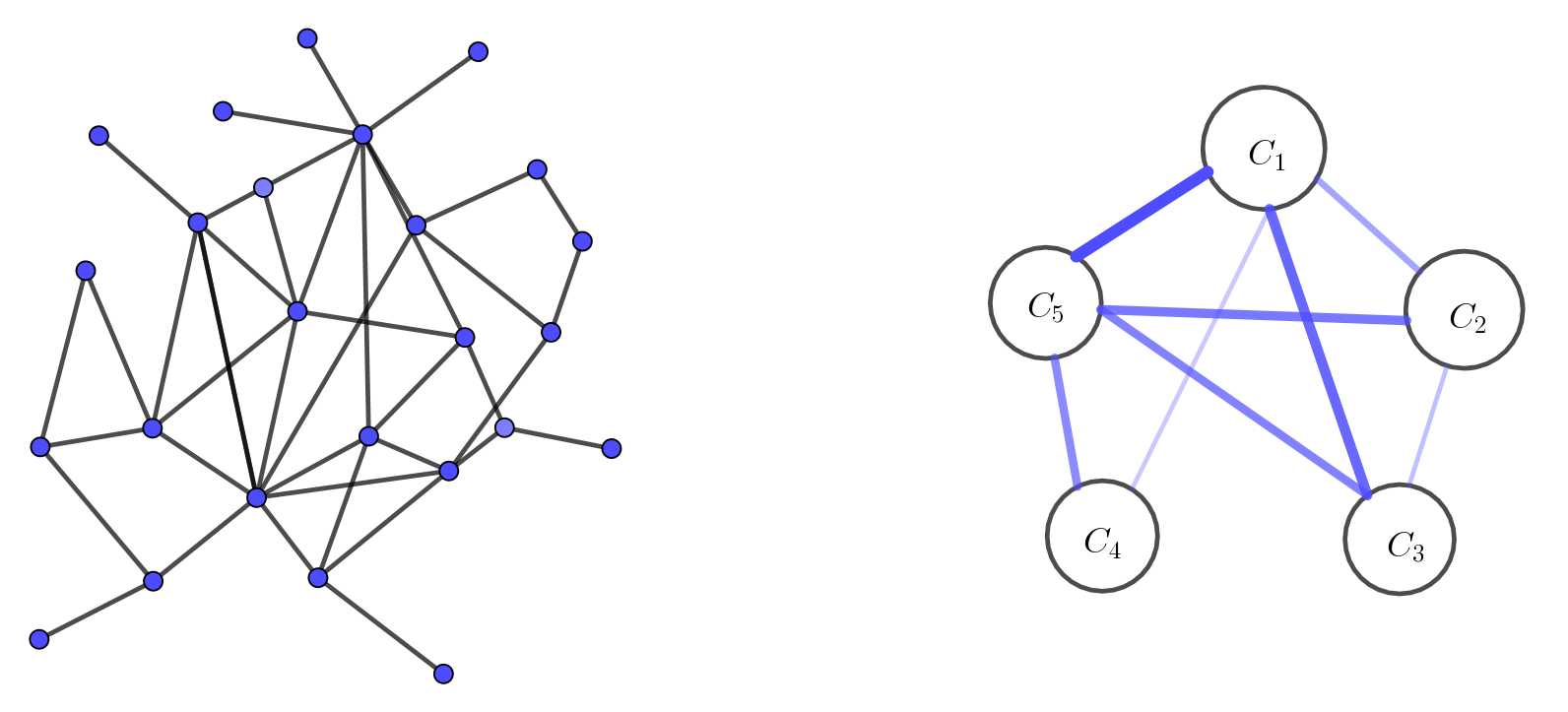}
\caption{\label{fig:approximation}Toy visualization of the Regularity Lemma, at left we have the graph and at right we have the approximation, with eight $\epsilon$-regular classes pair, the density of the pair is expressed with the color of the edges. Note: there could be some classes which are not regular (no edge).}
\end{figure}

To quantify the ``almost every'' statement we can say that there could be some pairs which are not ``quasi-random'' and which we can say nothing about, but the fraction of those will only be $\epsilon\binom{k}{2}$ as well. Another important thing is that each classes pair has its own density, Figure \ref{fig:approximation} visualize the concept

\paragraph{Technical Version} Now we can enter into the technical details of the Lemma. First we introduce the notion of density of a classes pair. Let $G = (V,E)$ be an undirected graph with no self-loops, and let $X,Y \subseteq V$ be two disjoint subsets of vertices of $G$. 

\begin{definition}{(Pair Density)}\label{def:density}
We define the edge density of the pair $(X,Y)$ as:
$$d(X,Y) = \frac{e(X,Y)}{|X||Y|}$$
\end{definition}  

This is just the amount of edges shared by the two sets divided by the maximum number of edges that there could be between the two sets, naturally the quantity lies between 0 and 1.

Now we are going to define what it means to a pair to be ``quasi-random'', in particular we will call the pair $\epsilon$-regular and the property is defined as follows:

\begin{definition}{($\epsilon$-regularity)} A pair $(A,B)$ of sets in $V(G)$ are $\epsilon$-regular if $\forall X \subset A, Y \subset B$ such that $|X| \geq \epsilon |A|, |Y| \geq \epsilon |B| $ we have:
$$ \left | d(X,Y) - d(A,B) \right | < \epsilon$$
\end{definition}

Which translated says: given a pair of vertices sets $A,B$, we say they are $\epsilon$-regular if by taking two sufficiently small subsets of the two the density between them is roughly the same as the density of $A,B$. This indicates some sort of homogeneity of the structure of the pairs.\\ 

In order to define Szemerédi's Regularity Lemma we last need another tool which is the Embedding Lemma (also known as Key Lemma) introduced by Komlos et al. \cite{Komlos1996SzemeredisTheory} \cite{Komlos2002TheTheory}, that actually represents the crucial property of $\epsilon$-regularity. Before stating the lemma we need to clarify some notation: Given a graph $H$ and an integer $m$, let $H(m)$ denote the graph obtained by ``blowing up'' each vertex of $H$ to size $m$, i.e., each vertex $j \in V(H)$ is replaced by a set $A_{j}$ of size $m$. Thus $H(m)$ has vertex set $\bigcup_j A_j$, and edge set $\{uv:u \in A_j$ for some $ij \in E(H)\}$. 

Given a graph $H$, an integer $m$ and $\delta >\epsilon>0$, let $\mathcal{G}(H,m,\epsilon,\delta)$ denote the family of graphs $G$ such that $V(G) = V(H(m)), G \subset H(m)$, and $G[A_i, A_j]$ is $\epsilon$-regular and has density at least $\delta$ whenever $ij \in E(H)$.

\begin{Lemma}{(Embedding Lemma)} 
For every $\epsilon>0, \delta>0$ there exist a $n_0$ such that if $V(G)= A \cup B \cup C, $
Let $\Delta \in \mathbb{N}$, let $\delta > 0$, and let $\epsilon_0 = \delta^\Delta / (\Delta+2)$. Let $R$ be a graph, let $m,t \in \mathbb{N}$ with $t \leq \epsilon_0m$, and let $H \subset R(t)$ with maximum degree at most $\Delta$. If $\epsilon_0 > \epsilon > 0$ and $G \in  \mathcal{G}(R, m, \epsilon, \delta+\epsilon)$, then $G$ contains at least $(\epsilon_0m)^{V(H)}$ copies of H.
\end{Lemma}

Which translated says that for the purpose of finding small subgraphs in a graph $G$, we can treat sufficiently dense $\epsilon$-regular pairs like complete bipartite graphs. A consequence of the latter fact is that the graph that we want to approximate should have a fairly high density, it is very important to keep this in mind because the lemma does not apply for very sparse graphs. Ultimately it states that the topological structure of the graph which approximates the original one, is guaranteed to be present inside the original one at least a certain number of times. This lemma guarantees the fact that we are able to compress/approximate a graph with a regular partition and still being able to preserve topological structures. Now we are ready to fully states the regularity lemma.

\begin{Lemma}{(Szemerédi's Regularity Lemma)} Let $\epsilon > 0, m \in \mathbb{N}$. There exists $k_0=k_0(\epsilon,m) \in \mathbb{N}$ such that for any graph $G$ there exists a partition\footnote{sometimes also called equitable partition} $V(G)=C_0 \cup C_1 \cup \dots \cup C_k$ where $m \leq k \leq k_0$ such that:

\begin{itemize}
\item $|C_0| \leq \epsilon |V(G)|$
\item $|C_i| = \dots = |C_k|$ 
\item $(C_i, C_j)$ is $\epsilon$-regular, for all but $\epsilon\binom{k}{2}$ pairs ${i,j}$ with $1 \leq i < j \leq k$
\end{itemize}
\end{Lemma}

\paragraph{Notes}All the classes must have the same cardinality except for the set $C_0$ that can have less nodes inside. As already pointed out there could be some irregular pairs, the fraction admitted can not exceed $\epsilon\binom{k}{2}$. Another very important requirement, has been already mentioned: the analyzed graph should be sufficiently dense, it is a direct consequence of the Embedding Lemma. We also absolutely need to report this important theorem from Gowers \cite{Gowers1997}:

\begin{theorem}{(Gowers 1997)}
For any  $\epsilon > 0$, there is a graph so that any application of the Regulrity Lemma requires that the number of clusters is at least a number which is a tower of twos of height proportional to $\log{\frac{1}{\epsilon}}$.
\end{theorem}

It says that the smaller the $\epsilon$ parameter (the approximation factor) the more clusters (classes) we need to have in a $\epsilon$-regular partition, in fact we should have a huge number of classes since there is a tower constant to be respected. This implies that the original graph size is astronomically big and this constitutes the major factor which makes the implementation infeasible.

Said that and with these assumptions in mind, we now move to the algorithmic formulation.

\section{Alon et al.'s Algorithm}\label{sec:alon}

In their original paper Alon et al. \cite{Alon94} showed that the complexity of deciding if a given partition is $\epsilon$-regular is more difficult than actually building one then they devised an algorithm to build one. In fact they showed that creating an $\epsilon$-regular partition can be done with a feasible temporal complexity $O(n^2)$. Let's dig into the algorithm: we will report only the main theorem and some definitions which are crucial to understand the structure of the algorithm.\\ 

The following theorem is the core of Alon et al.'s algorithm, take note that we are not going to prove any theorem since all the proofs can be read in the original paper and are all quite long to handle.

\begin{theorem}{(Alon et al., 1994)}
For every $\epsilon > 0$ and every positive integer $t$ there is an integer $Q = Q(\epsilon, t)$ such that every graph with $n > Q$ vertices has an $\epsilon$-regular partition into $k + 1$ classes, where $t \leq k \leq Q$. For every fixed $\epsilon > 0$ and $t \geq 1$ such a partition can be found in $O(M(n))$ sequential time, where $M(n) = O(n^{2.376})$ is the time for multiplying two $n \times n$ matrices with 0,1 entries over the integers. It can also be found in time $O(logn)$ on an EREW PRAM with a polynomial number of parallel processors.
\end{theorem}

As we already said, the theorem simply states that it exists an algorithm to build such partition and it has a feasible temporal complexity (but keep in mind that we still have the tower constant from Gowers on the number of classes required by a partition). The next question now is: How can we do this? Before moving on we need several other definitions. \\

Let $H$ be a bipartite graph with equal color classes $|A| = |B| = n$, from now on it will represent one of the pair of the final partition. We define the average degree of $H$ as:

$$\bar{d}(A,B)=\frac{1}{2n} \sum_{i \in A \cup B} deg(i)$$

where $deg(i)$ is the degree of vertex $i$.

For two distinct vertices $y_1, y_2 \in B$ define the neighbourhood deviation of $y_1$ and $y_2$ with:

$$\sigma(y_1, y_2) = |N(y_1) \cap N(y_2)| - \frac{\bar{d}^2}{n}$$

where $N(x)$ is the set of nighbours of vertex $x$. For a subset $Y \subseteq B$ the deviation of $Y$ is defined as:

$$\sigma(Y)=\frac{\sum_{y_1,y,2 \in Y}\sigma(y_1,y_2)}{|Y|^2}$$

Let $0 < \epsilon < 1/16$. It can be shown that if there exists $Y \subseteq B, |Y| > \epsilon n$ such that $\sigma(Y) \geq \epsilon^3 \frac{n}{2}$, then at least one of the following cases occurs:

\begin{enumerate}
\item $\bar{d} <\epsilon^3 n$ ($H$ is $\epsilon$-regular)
\item there exists in $B$ a set of more than $\frac{1}{8}\epsilon^4n$ vertices whose degree deivate from $\bar{d}$ by at least $\epsilon^4n$
\item there are subsets $A' \subset A, B' \subset B, |A'| \geq \frac{\epsilon^4}{n}n, |B'| \geq \frac{\epsilon^4}{n}n$ and $|\bar{d}(A', B') - \bar{d}(A,B)|\geq \epsilon^4$\label{alon3}
\end{enumerate}

In case the first two points does not hold one must find $A',B'$ sets as described in point \ref{alon3}. Now the proof of the latter Lemma provides an intuitive way to find the given subsets but, as we said, it is not reported here. Anyway, the main idea is that the procedure selects a subset of the nodes whose degree ``deviate'' the most from the average. More formally: for each $y_0 \in B$ with $|deg(y_0)-\bar{d}| < \epsilon^4n$ we find the set of vertices $B_{y_0} = \{y \in B| \sigma(y_0, y) \geq 2\epsilon^4 n \}$. The proof given in the original paper guarantees the existence of at least one such $y_0$ for which $|B_{y_0}| \geq \frac{\epsilon^4}{4}n$. The subsets $B'=B_{y_0}$ and $A'=N(y_0)$ are the required ones. These two subsets represent the collection of nodes that contribute more to the irregularity of the pair $(A,B)$, in fact they play an important role and we will address them as certificates. We will call complements all the other nodes of the class which are not part of the certificates. 

Alon et al.'s algorithm allows us to find the certificates which contribute more to the irregularity of a pair, before stating the algorithm we will report an important result of Szemerédi which combined with Alon et el.'s algorithm allows us to build such partitions. Given an $\epsilon$-regular partition $\mathcal{P}$ of a graph $G=(V,E)$ into $C_0,C_1,\dots, C_k$ classes Szemerédi \cite{Szemeredi75Graphs} defines a measure called index of partition:

$$sze\_ind(\mathcal{P})=\frac{1}{k^2}\sum_{s=1}^{k}\sum_{t=s+1}^{k}d(C_s,C_t)^2$$

Since $0\leq d(C_s,C_t) \leq 1, 1 \leq s, t \leq k$, it can be seen that $sze\_ind(\mathcal{P}) \leq \frac{1}{2}$

In his main proof, Szemerédi state that if a partition $\mathcal{P}$ violates the regularity condition, then it can be refined by a new partition and, in this case, the latter measure should increase. 

\paragraph{Alon et al.'s Alogirhtm}Now we are ready to see the main algorithm:

\begin{enumerate}
\item Create the initial partition: Arbitrarily divide the vertices of $G$ into an equitable partition $\mathcal{P}_1$ with classes $C_0,C_1, \dots, C_b$ where $|C_i|=\lfloor \frac{n}{b}\rfloor$ hence $|C_0|<b$
\item Check Regularity: For every pair $(C_r,C_s)$ of $\mathcal{P}_i$, verify if it is $\epsilon$-regular or find $X \subseteq C_r, Y \subseteq C_s, |X| \geq \frac{\epsilon^4}{16}C_1,|Y| \geq \frac{\epsilon^4}{16}C_1 $ such that $|\bar{d}(X,Y)-\bar{d}(C_s, C_t)| \geq \epsilon^4$
\item Count regular pairs: if there are at most $\epsilon \binom{k_i}{2}$ pairs that are not $\epsilon$-regular then, stop. $\mathcal{P}_i$ is an $\epsilon$-regular partition
\item Refine: Apply the refinement algorithm and obtain a partition $\mathcal{P}'$ with $1+k_i4^{k_i}$ classes
\item Go to step 2.
\end{enumerate}

Besides all the notation and mathematics here we have an algorithm which guarantees the possibility to reach an $\epsilon$-regular partition that respects the strong regularity conditions. Several other algorithms have been created most of them relaxed the conditions of $\epsilon$-regularity to a weak notion, in particular the Frieze-Kannan algorithm \cite{Frieze1999APartition} is the most famous one, which is based on an intriguing relation between the regularity conditions and the singular values of matrices.

%% file: devisedtechnique.tex

\section{Overview of the CoDec Procedure}\label{sec:codec}

\begin{figure}
\centering
\includegraphics[width=0.65\textwidth]{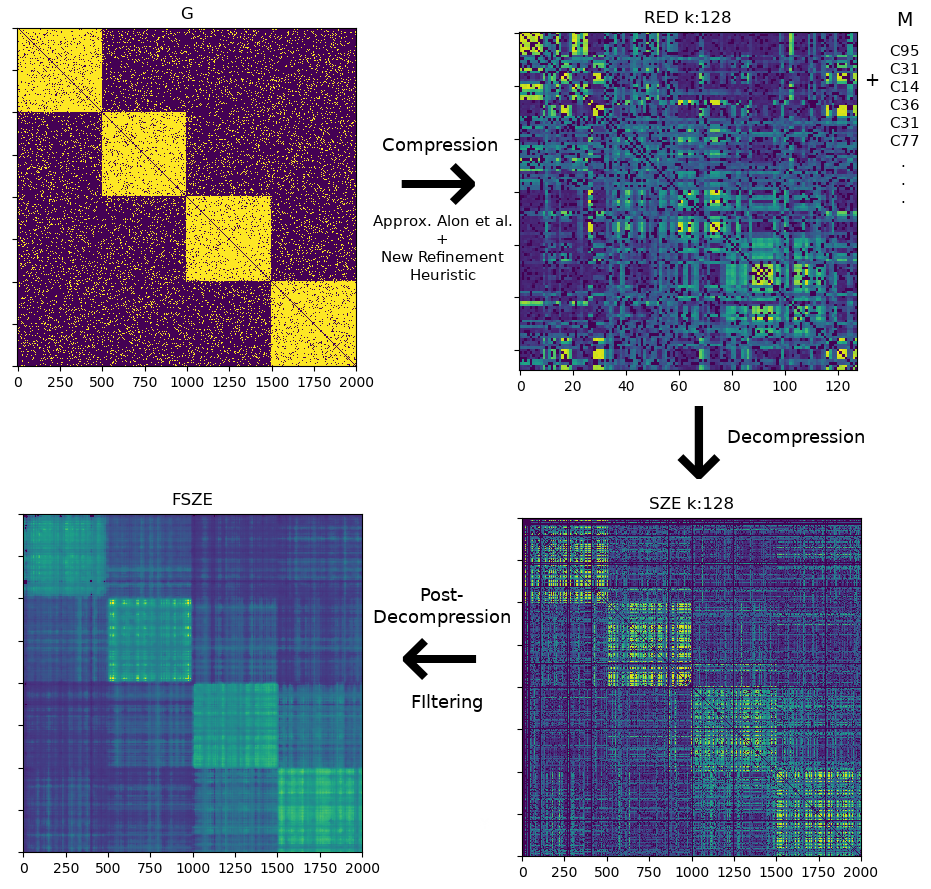}
\caption{\label{fig:codec}Schematic representation of the CoDec Procedure. The input graph is a noisy graph with four balanced clusters representing the structure carrying information. The compression step produces a partition of $128$ classes coded in the $RED$ matrix plus a vector of membership $M$. The second phase exploits the Embedding Lemma to reconstruct the compressed structure. The third and last step Post-Decompression applies a filter to highlight the structural preservation.}
\end{figure}

Thanks to the previous sections we now have a basic context and idea of the domain of the devised technique. In particular, now we have a grasp on how Alon et al's algorithm works, therefore we can describe the CoDec Procedure (Co-mpression Dec-ompression) which is the core of the work done during this thesis. We will split the algorithm in different steps as Figure \ref{fig:codec} shows, but before describing the work, let's just clarify some aspects that are valid throughout all the description.
\paragraph{Notes}
\begin{itemize}
\item Since we are entering to a more algorithmic description of the work, from now on we will use matrices as data structures to represent any undirected graph with no self loops. In particular the matrices are symmetric with an empty diagonal (no self loops).
\item From now on with the words ``reduced graph'' or ``compression'' we will refer to a $\epsilon$-regular partition $\mathcal{P}$ (of size $k$) of an undirected graph with no self loops achieved by running the approximated version of Alon et al.'s algorithm in addition with the new heuristic. We sometimes address it with the symbol $\mathcal{P}$ or in its matrix representation with $RED$ (from REDuced graph).
\end{itemize}

\paragraph{Description}
The procedure takes as input an undirected (weighted or unweighted) graph of $n$ nodes, expressed as a symmetric $n \times n$ matrix $G$, then:

\begin{enumerate}
\item It searches and collect different $\epsilon$-regular partitions $\mathcal{P}_1, \mathcal{P}_2 \dots $ by running several approximated Alon et al.'s algorithm instances in conjunction with the new refinement heuristic, over a set of $\epsilon$ candidates. In particular, the procedure uniformly sample the $\epsilon$ parameter space in the interval $[0,0.5]$. \label{main:search}

\item[1b.] We select the best partition $\mathcal{P^*}$ from the collected ones.
The partition is the one which both maximizes the partition cardinality $|\mathcal{P}|$ (therefore maximizes the Szemerédi Index according to its formulation) and minimizes the value of the parameter $\epsilon$ (the resulted graph should be a better approximation of the graph according to the theory). We then build a $|\mathcal{P^*}| \times |\mathcal{P^*}|$ matrix to contain the compression of the matrix $G$ that we will call $RED$.\label{main:compressed}

 \item The procedure decompress $RED$ by exploiting the Embedding Lemma (Section \ref{sec:decompression}), giving us a weighted $n \times n$ matrix that we will call $SZE$ and which should carry the structural properties of the graph.

\item We then apply a filtering $f(\cdot)$ technique to the matrix to highlight the patterns in the $SZE$ matrix giving us the final $n \times n$ matrix $f(SZE)$ or $FSZE$. This procedure is described in Section \ref{sec:filtering}.
\end{enumerate}

\begin{algorithm} \label{pseu:codec}
	\caption{CoDec Procedure}
	\begin{algorithmic}[1]
    	\Procedure{codec}{$G$}
        \For{$\epsilon$ in $(0,1)$ with step of $0.1$}
        	\State $\mathcal{P}_i =$ \Call{approx\_alon}{$G, \epsilon$}
			\State Store $\mathcal{P}_i$ in a vector $V$
		\EndFor
        \State $\mathcal{P}^* =$ \Call{best\_partition}{$V$}
        \State $RED,M$ = \Call{compress}{$\mathcal{P}^*$} 
        \State $SZE =$ \Call{decompression}{$RED,M$}
        \State $FSZE =$ \Call{post-decompression}{$SZE$}
		\EndProcedure
	\end{algorithmic}
\end{algorithm}

\paragraph{Observations}
In point \ref{main:search} a smarter approach has been considered but not tested yet, it consists in a binary search with a parametrized step size over the $\epsilon$ parameter space $[0,0.5]$. This improvement would significantly reduce the number of runs and speed up the overall running time, although the required time for a single run in a medium-size graph ($n<10000$) is acceptable. This is possible due to the fact that, experimentally, it has been noted that the cardinality of the partitions found grows as $\epsilon$ grows until we are no more able to find any useful partition.

The compression in point \ref{main:compressed} is expressed with a mapping from the nodes to their membership to a particular class trough a $1 \times n$ vector, which encodes the nodes class membership $M$, plus a $|\mathcal{P^*}| \times |\mathcal{P^*}|$ matrix $RED$ to encode the densities between the classes\footnote{We use only half of the matrix achieving a $O(|\mathcal{P}|^2)$ space complexity with $\frac{1}{2}$ constant factor}.

\section{A New Refinement}\label{sec:refinement}
In this Section we see the main idea that this work proposes as heuristic for the refinement procedure of Alon et al.'s approximated algorithm. 

As we said, in the original paper of Alon et al. they show that the Regularity Lemma can be formulated in algorithmic terms, but they do not provide a constructive version of how to reach the desired partition through the refinement step. Other works tried to implement some kind of heuristics \cite{Neil2014ExperimentalClustering} but the choices are pretty simple and only take in consideration properties of the nodes. With this thesis we introduce a new heuristic which exploits more grounded intuitions.
 
\begin{figure}
\centering
\includegraphics[width=0.9\textwidth]{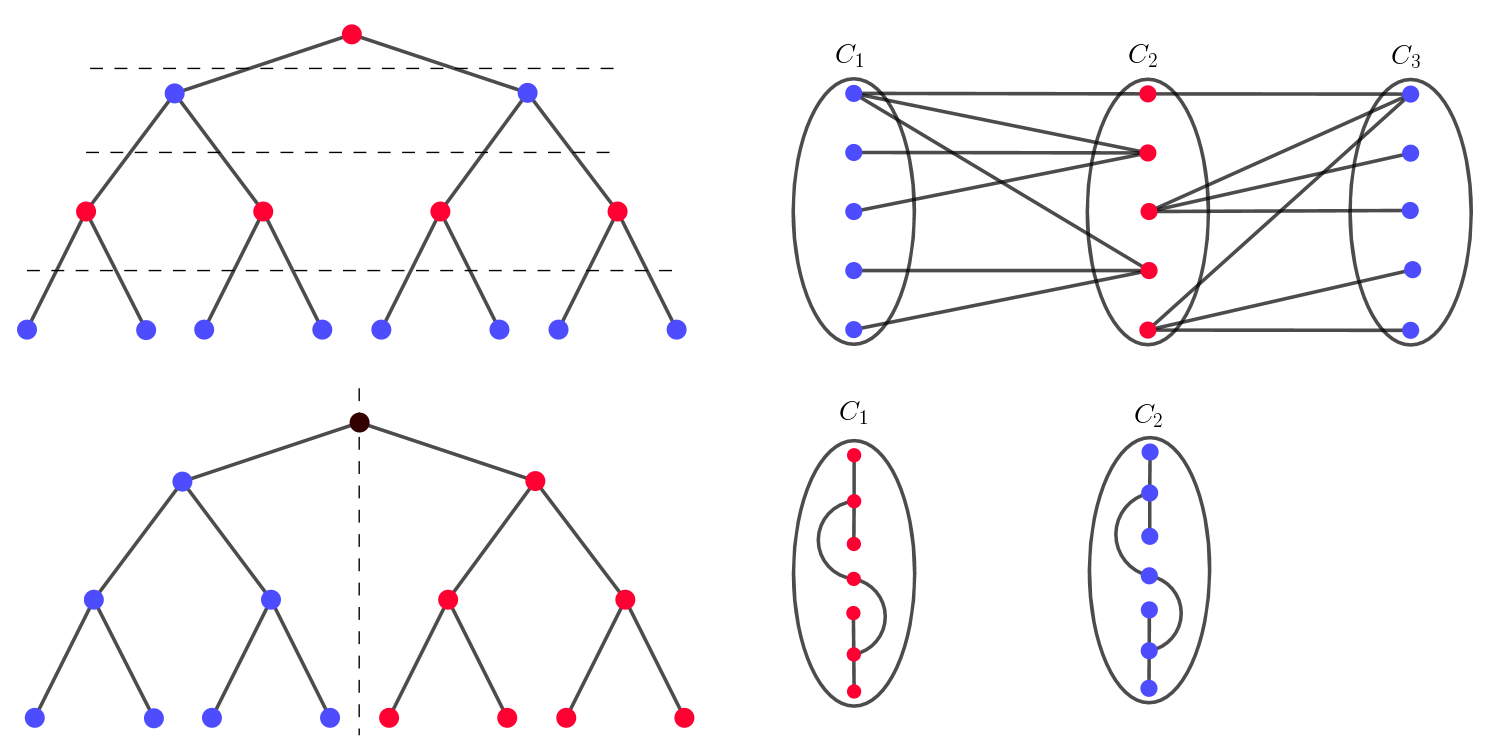}
\caption{\label{fig:policy}A toy example: it can be seen that by achieving different indegree average values, we converge to a particular coloring of the graph. In the upper image we have an internal average degree of $0$ while in the case below we have an internal average degree of $1.81$ and in fact they represent two different colorings.}
\end{figure}

\paragraph{Idea}
The key idea of this refinement is that we now try to reach the final partition by exploiting properties of the classes instead of the nodes. In particular it can be drawn a correspondence between the average internal degree of the nodes in a class and a particular graph coloring. As can be seen in Fig. \ref{fig:policy} when we drive the internal degree\footnote{By internal degree we mean the degree of a node with respect to the nodes in the same class} of the nodes inside a class, we are coloring the graph by highlighting some regular patterns. We then assume that the information to be preserved is represented in the repetition of some sub-patterns and we drive the refinement to keep these substructures between the classes. By trusting the Alon algorithm intuition we then aim to converge to a meaningful partition which highlights the structure of the input graph. Obviously the analyzed graph must have some intrinsic patterns in order for this idea to work.

\paragraph{Details} We need to specify some details and decisions made to fully understand the heuristic:
\begin{itemize}
\item At the very beginning we need to provide a starting partition for the approximated Alon et al.'s algorithm, the choice is to split the graph in 4 classes and distribute the nodes in a random way.
\item In the original definition the number of classes at each iteration should grow astronomically with a tower function as lower bound. Since this detail is computationally infeasible we decided that the number of classes always doubles at each iteration.
\end{itemize} 
\paragraph{Notations} We need to clarify the notation since there are several entities to be considered: 

\begin{itemize}
\item $X_i, X_j$ : certificates of the irregular pair $C_i$ $C_j$
\item $\bar{X_i},\bar{X_j}$ : complements of the irregular pair $C_i$ $C_j$.  Just to be clear: $X_i \cup \bar{X_i} = C_i$
\item $d(C_i, C_j)$ : density between the two classes as specified in Section \ref{def:density}. If $i=j$ then we consider the internal density (or in-density) defined as 

$$d(C_i,C_i)=\frac{e(C_i,C_i)}{|C_i|^2}$$

\item $C_i^1, C_i^2$: notation to express the two new classes created from an irregular pair \footnote{just for clarity there must be $C^{j+1}, C^{j+2}$ too, but for brevity we just consider the class $i$ over the whole description}
\item indegree : degree of a node calculated with respect to the nodes of the same class
\item in-hubs : hub nodes with respect to the indegree i.e. nodes with highest indegree value
\end{itemize}

\paragraph{Description}We are now ready to define the new refinement procedure. Given a partition $\mathcal{P}_i= C_0, C_1, \dots, C_k$ at some iteration $i$ of approximated Alon et al.'s  algorithm:
\begin{enumerate}
\item Randomly pick a class $C_i$ 
\item \label{refinement:score} If $C_i$ is irregular with other classes then choose the candidate $C_j$ which maximizes the following score:
$$ S = d(C_i, C_j) + (1 - |d(C_i, C_i) - d(C_j, C_j)|) \rightarrow [0,2]$$
Now we have a classes pair candidate $(C_i, C_j)$ ready to be refined\footnote{each class will be split in two, giving birth to 4 new classes $C_i^1, C_i^2,C_j^1,C_j^2$}. We then process the two certificates of the pair, in particular we have two possible scenarios for each certificate. We consider only $X_i$ for the sake of illustration, but the exact same steps will be applied for $X_j$:
  \begin{enumerate}
  
  \item Case $d(X_i,X_i) \geq 0.5$ (Densification)\label{refinement:densification} 
    
    \begin{enumerate}
    \item Sort $X_i$ nodes by indegree to reveal in-hubs $\{h_1, h_2, \dots , h_{|X_i|}\}$  
    \item Unzip hubs to new classes : $C_i^1 : \{h_1, h_3, h_5, \dots , h_{|X_i|-1}\}$ and $C_i^2 : \{h_2, h_4, h_6, \dots , h_{|X_i|}\}$   
    \item Fill $C_i^1$ and $C_i^2$ up to $|C_i|/2$ by taking the ``most''\footnote{Weighted case: the nodes with the highest value. Unweighted case: just the nodes which share the most connections with the considered set} connected nodes from the union of the complements $\bar{X_i} \cup \bar{X_j}$  
    \end{enumerate}
  
  \item Case $d(X_i,X_i) < 0.5$ (Sparsification)\label{refinement:sparsification}
  	\begin{enumerate}
  	\item Random sampling $|X_i|/2$ nodes into $C_i^1$ and $C_i^2$
    \item Fill $C_i^1$ and $C_i^2$ up to $|C_i|/2$ by taking the ``least'' connected nodes from the union of the complements $\bar{X_i} \cup \bar{X_j}$
  	\end{enumerate}
  \end{enumerate}
  
  \item If class $C_i$ is $\epsilon$-regular with all the other classes
  	\begin{enumerate}
  	\item Sort $C_i$ nodes by indegree to reveal in-hubs $\{h_1, h_2, \dots , h_{|C_i|}\}$  
    \item Unzip nodes to new classes : $C_i^1 : \{h_1, h_3, h_5, \dots , h_{|C_i|-1}\}$ and $C_i^2 : \{h_2, h_4, h_6, \dots , h_{|C_i|}\}$
  	\end{enumerate}
    
  \item Repeat the above points until there are no more classes left.
  
  \item[*] Since the new refinement split each class in two, it could happen that we have an odd cardinality of a class, in this case we put the exceeding node in the ``trash-set'' $C_0$. When the cardinality of this set exceeds $\epsilon n$ and we have enough nodes, we equally distribute them between the classes, otherwise the procedure declares the partition irregular.
\end{enumerate}

\begin{algorithm} \label{pseu:newrefinement}
	\caption{Refinement step performed at the $i$-th iteration of the approximated Alon et. al's algorithm}
	\begin{algorithmic}[1]
		\Procedure{refinement}{$\mathcal{P}^i$}
			\For {each class $C_i$ in $\mathcal{P}^i$}
				\If{$C_i$ is $\epsilon$-regular with all the other classes}
					\State $C_i =$ \Call{sort\_by\_indegree}{$C_i$}
					\State $C^1_i,C^2_i = $\Call{unzip}{$C_i$}
                    \Comment This statement may add a node to $C_0$ if $|C_i|$ is odd
                \Else
					\State Select $C_j$ with most similar internal structure   
					\State Get certificates $(X_i, X_j)$ and complements $(\bar{X_i},\bar{X_j})$ of the pair
					\If{$d(X_i,X_i)<0.5$}
                    	\State $C^1_i,C^2_i =$ \Call{sparsification}{$X_i,\bar{X_i} \cup \bar{X_j}$}
                    \Comment This statement may add a node to $C_0$
                        
                    \Else
                    	\State $C^1_i,C^2_i =$ \Call{densification}{$X_i,\bar{X_i} \cup \bar{X_j}$}
                    \Comment This statement may add a node to $C_0$
                        
                    \EndIf
					\State Perform step 9,10,11,12 for $X_j$
				\EndIf
			\EndFor
			\If{$|C_0| > \epsilon n$ and $|C_0|>|\mathcal{P}^{i+1}|$}
            	\State Uniformly distribute nodes of $C_0$ between all the classes
            \Else
            	\State \Return $(\mathcal{P}^{i+1}, irregular)$
            \EndIf
            \State \Return $(\mathcal{P}^{i+1}, regular)$
		\EndProcedure
	\end{algorithmic}
\end{algorithm}

\paragraph{Observations} Since this new refinement is a heuristic we don't have any formal proof of the correctness, but we can give some intuitive explanations of the procedure.

The point \ref{refinement:score} defines a score in order to select the irregular pair $C_i, C_j$ that we will refine (split in $C_i^1, C_i^2$ and $C_j^1, C_j^2$) in the next step. In particular this score aims to select pairs who share the most similar internal structure, this is to follow the intuition of the heuristic which has been illustrated with Figure \ref{fig:policy}. By selecting the most similar classes pair we aim to select a particular topological substructure, at least this should be the connection. After that we proceed to purify the two classes with step \ref{refinement:densification} and \ref{refinement:sparsification} in other words we aim to converge to a pattern.

The Densification branch \ref{refinement:densification} also contribute to maximize the Szemerédi index since we are unzipping in-hub nodes to the new two classes, therefore this should result in a higher density between the new classes and a higher Szemerédi index (recall: the Szemerédi Index grows if we have a very dense graph or if we have a huge amount of classes).


\section{Graph Decompression}\label{sec:decompression}
In this Section we describe the decompression phase where we take the structure that represents the best partition and we decompress it to have a weighted matrix of the same size of the original one. In this particular phase we fully exploit the Embedding Lemma that, as we said, is the reason why the Regularity Lemma is useful. At this point we have selected the best partition $\mathcal{P}^*$ coded through a $1 \times n$ vector which represents a mapping of the nodes to their respective class, plus a $|\mathcal{P^*}| \times |\mathcal{P^*}|$ matrix $RED$ to encode the densities between the classes.

The procedure decompresses the matrix by fully connecting the nodes of each $\epsilon$-regular classes pair. It is mandatory to fully connect the nodes because that's a requirement of the Embedding Lemma as we already described in Section \ref{sec:szemeredilemma}. In particular given an $\epsilon$-regular pair with density $d$ we will fully connect the nodes by assigning a weight equal to $d$. This means that the decompression phase generates a weighted undirected graph even if the input graph is unweighted. The resulting matrix encodes the structural patterns of the original graph, ideally higher the density between the pair the more structure we captured. The latter consideration is a consequence of the fact that higher density pairs contribute the most to the partition index, therefore it means that a pair with high density should contain more ``structural information''.
We will call the reconstructed/decompressed $n \times n$ weighted matrix $SZE$.

\begin{figure}
\centering
\includegraphics[width=0.8\textwidth]{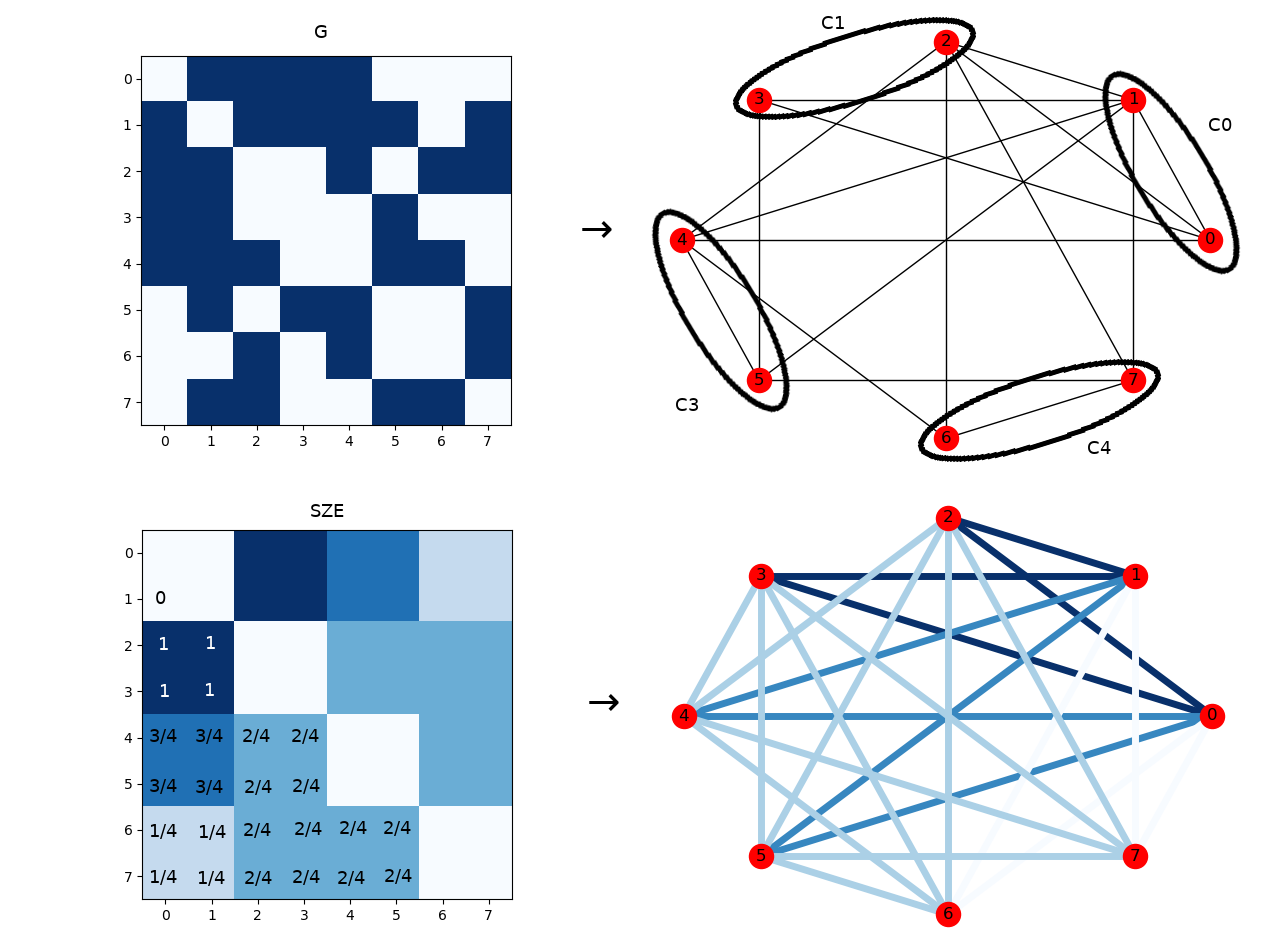}
\caption{\label{fig:example}Consider the partition pair $(C_3,C_4)$ its density is $\frac{1}{2}$ since there could be 4 edges between them but in $G$, which is our corrupted graph, we have only 2 edges between them. In the $SZE$ reconstruction we will fully connect the nodes inside the classes by setting the edges weights equal to the density between the pair. From this toy example we could also see how the Szemerédi's Regularity Lemma can be applied as a tool to approximate the corrupted graph $G$ and to reveal the structural pattern of the original graph (let's assume that the original graph was a fully connected graph).}
\end{figure}

The reconstruction used in the experiments is very simple and it is label-preserving, this is a very important feature because allows us to visualize the results and compute some error measures. In order to follow the theory, the reconstruction phase does not put edges between nodes of the same class, or between the $\epsilon$-irregular pair, although the framework provides such possibility. In particular if we also connect the irregular pairs the result would be a fully connected approximated graph, experimentally it has been noted that this feature does not significantly improve the results.

Another decompression feature has been taken in consideration: we can try to keep track also of the internal densities of the classes. The idea is to exploit this information when we blow up the reduced matrix. In particular we could create an Erd\"{o}s-Rényi graph with probability equal to the internal density of the class in order to approximate its internal structure. Unfortunately this feature has not been extensively studied, so the results has been omitted.

\section{Post-Decompression filtering}\label{sec:filtering}

In this phase we abstract the notion of the graph and we treat its data representation i.e. the matrix purely as an image in order to perform some morphological operations. This turns out to be a good trick to mitigate the approximation of our algorithm and to let the structural pattern emerge from the decompressed matrix.

\begin{figure}
\centering
\includegraphics[width=0.6\textwidth]{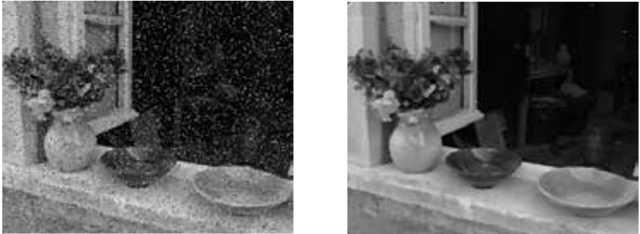}
\caption{\label{fig:median}An example of median filtering, taken from Szeliski \cite{Szeliski2010ComputerApplications} book.}
\end{figure}

In particular after the Alon compression and reconstruction, we filter the matrix $SZE$ with a fixed size median kernel giving birth to the $FSZE$ matrix. Median filtering is a nonlinear morphological operation mainly used in Computer Vision since it is very effective at removing ``salt and pepper'' noise in images Figure \ref{fig:median} illustrate such application. It works by moving a kernel through the image pixel by pixel, replacing each value 
with the median value of neighbouring pixels.

The pattern of neighbours is called the ``kernel'' (or window) and it slides over the entire image. The median is calculated by first sorting all the pixel values from the window into numerical order, and then replacing the pixel being considered with the middle (median) pixel value (50 percentile). The filtering phase is essential to clean up the reconstructed matrix since in the reconstruction phase we fully connects the classes and obviously we could add some extra edges. In particular, as we mentioned in the decompression section, the edges with higher weight should carry more structural information, to filter the noise of the low density weights we run a median kernel that should highlight the matrix structural properties. 

Experimentally we decided to use a fixed window size, but in principle one can study the optimal size in function of some graph invariant although this possibility has not been taken in consideration since a fixed size provided a good solution.

When we apply any filter in images there is a problem when taking in account the pixels in the border of the image, in our case we adopted a reflection padding technique to handle matrix borders. There is not a strong motivation to use a reflection padding, it is purely an implementation detail since it does not severely affect the result.

%% file: experimentalresults.tex
In this section we report the experimental setup, the results obtained and we will draw some considerations. First of all let's clarify that two different experiments have been conducted. The first one reported in Section \ref{sec:synthexperiments} studies the CoDec Procedure applied to various Synthetic Graphs, the objective is to do a deep study of the devised technique. The second type of experiments reported in Section \ref{sec:realexperiments} uses Real Networks, the aim is to verify the procedure over real cases and to assert the approximation feature of the framework. 

Almost all the experiments involving medium size graphs ($n<5000$ nodes), have been conducted with my personal machine: an Intel Core i5 @ 2.60GHz HP Pavilion 15 Notebook with 8GB of RAM (DDR3 Synchronous 1600 MHz) running Arch-Linux with kernel version 4.14.4-1, but the experiments involving graphs with $n\geq 5000$ have been conducted in the \href{https://www.dais.unive.it/scscf/it/scscf/}{SCSCF} which is the scientific computational cluster made of 7 nodes of the Ca' Foscari university. The reason for the latter choice is for both time (since we performed a lot of repetitions) and resources reasons (my personal Notebook we are able to process single runs of graphs up to 17000 nodes, but it couldn't work all day for several days without causing damage to the hardware).

\section{Evaluation measures}\label{sec:measures}

In order to validate the results, in the experimental evaluation we have used three main measures. The first two are somehow a ``direct'' estimation of dissimilarity between any two matrices, to evaluate the approximation quality of the method. As we will discuss we could argue if these two provide a good estimation. In fact we will discuss it later on the consideration section. The last measure estimates the structural preservation of the technique by comparing two clustering vectors. 

\subsection*{Direct Measures}
Two direct measures have been used to compute the dissimilarity between the original graph and the result of our procedure. To illustrate the measures let's define two matrices $\mathbf{A} = (a_{ij})$ and $\mathbf{B} = (b_{ij})$ both of size $n \times n$:
\begin{itemize}
\item Normalized $l_1$-distance
$$l_1(\mathbf{A}, \mathbf{B}) = \frac{\sum_{i=1}^n \sum_{j=1}^n |a_{ij} - b_{ij}|}{n} \rightarrow [0,1]$$
\item Normalized $l_2$-distance
$$l_2(\mathbf{A}, \mathbf{B}) = \frac{\sqrt{\sum_{i=1}^n \sum_{j=1}^n (a_{ij} - b_{ij})^2}}{n} \rightarrow [0,1]$$
A high value of one of the two measure means that the two matrices are very ``distant'' or dissimilar, while lower values means similar matrices. We would like to have low values.
\end{itemize}

\subsection*{Structural Preservation Measure}
In order to measure the preservation of the structure in cluster-like graphs, it has been implemented a column-wise Knn Voting System clustering technique (we will call it KVS). Given a graph, the KVS produces a clustering vector, to validate the correctness of the grouping, then we calculate the Adjusted Rand Index ($ARI$) between the graph's original clustering $L$ and the predicted one $C$. 

\paragraph{KVS Description} KVS-clustering technique works as follows:
\begin{enumerate}
\item Given a weighted matrix, for each row of the matrix we take the column positions of the $k$ highest values
\item We check, in the these positions, which are the true clustering labels by looking at the  respective positions in $L$, giving us $k$ labels predictions
\item \label{kvs:kvalue} We set the current row label prediction to the label which is the most repeated in the $k$ predictions 
\end{enumerate}

Since the KVS-clustering requires a parameter $k$ to run, we try different values and then we pick the one which maximizes the ARI, specifically we tried with three different values of $k=5,7,9$ we must be sure that $k$ is an odd number to always have point \ref{kvs:kvalue} satisfied.

\paragraph{ARI Description} The KVS clustering technique outputs a $1 \times n$ array of predicted labels $C$, when we calculate the Adjusted Rand Index with respect to the labeling vector $L$, the similarity score is between -1 and 1. A random prediction clustering has an ARI close to 0 while 1 stands for perfect match, -1 means completely opposite labeling.

This estimation has been introduced by Hubert et al. \cite{Hubert1985} and it is a measure to compare two clusterings. If L is a ground truth clustering vector and $C$ is the clustering that we want to evaluate, let us define $a$ and $b$ as:
\begin{itemize}
\item $a$: the number of pairs of elements that are in the same set in $L$ and in the same set in $C$
\item $b$: the number of pairs of elements that are in different sets in $L$ and in different sets in $C$
\end{itemize}

The raw (unadjusted) Rand Index (RI) is then given by:
$$\text{RI} = \frac{a + b}{L_2^{n_{samples}}}$$
Where $L_2^{n_{samples}}$ is the total number of possible pairs in the dataset (without ordering).
However the RI score does not guarantee that random label assignments will get a value close to zero (especially if the number of clusters is in the same order of magnitude as the number of samples).
To counter this effect we can discount the expected Rand Index $E[\text{RI}]$ of random labelings by defining the adjusted Rand index as follows:
$$\text{ARI} = \frac{\text{RI} - E[\text{RI}]}{\max(\text{RI}) - E[\text{RI}]}$$

\section{Synthetic Graphs Experiments}\label{sec:synthexperiments}

\subsection*{Synthetic Graph Generation}\label{ssec:synthgeneration}

In this subsection we will see how the synthetic graphs, used for the experiments, has been generated. As already said, we are interested to undirected graphs that can be seen as a composition of ``structure'' which carries the knowledge of the phenomenon, plus some ``noise''.

That said, we can formalize a bit the notion of the graphs that we will analyze, in particular we can see them as:

$$G=GT+N$$

Where $G$ represents the graph, $GT$ (namely Ground Truth) represents the structure and $N$ represents the noise. 

Given a graph $G$ of $n$ nodes, we model the noise with Erd\"{o}s-Renyi Graph (ERG) of size $n$ with probability of connection between two nodes $p$ (from now on we will call this parameter inter-cluster noise level, or inter-noise). The structural part $GT$ is a graph of $n$ nodes with $C$ disconnected groups (cluster) of nodes of variable size.

\begin{figure}
\centering
\includegraphics[width=0.75\textwidth]{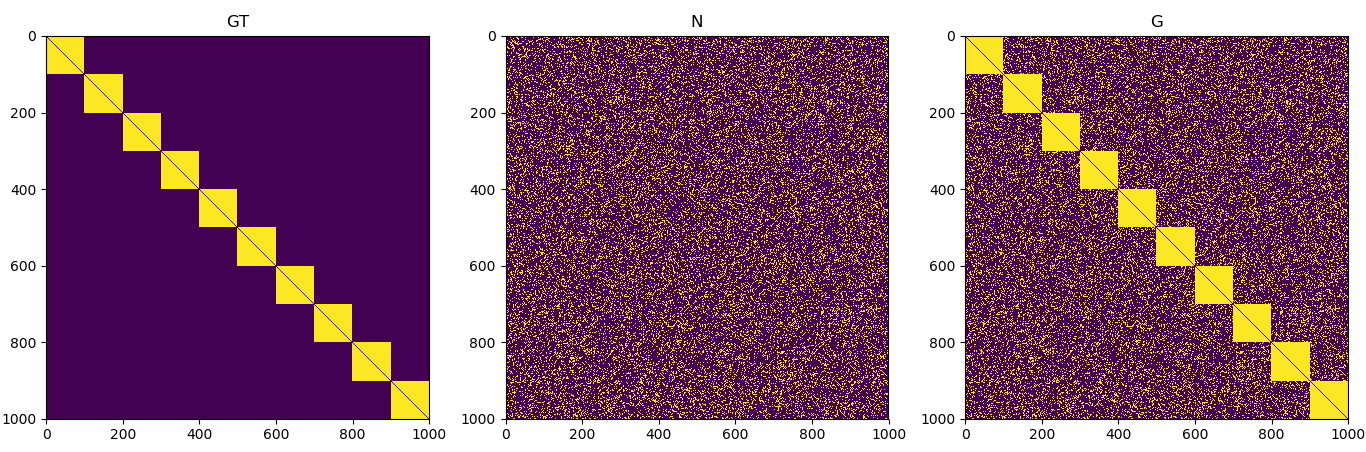}
\caption{\label{fig:synthg}A Synthetic Graph $G$ composed by 1000 nodes with internoise of 0.2 and 10 clusters. The labeling vector will be a $1 \times 1000$ vector where the first 10 elements will be set to 1 (first cluster label) then the next 10 elements will be set to 2 and so on to 10.}
\end{figure}

For the evaluation of the experimental results, we also need a vector of labels $L$ that represents the correct clustering of the nodes according to the ground truth. The labeling vector is a $1 \times n$ vector which indicates the membership of a node to the respective cluster. Figure \ref{fig:synthg} exemplifies the creation of a synthetic graph.

\paragraph{Basic Version} Now we describe the basic procedure to generate unweighted Synthetic Graphs: 
\begin{itemize}
  \item Generate a $n\times n$ adjacency matrix $A$ with $A_{ij}$ equal to a realization of a $U(0,1)$ r.v..
  \item If $A_{ij} \leq $ inter-noise level then put an edge (we're creating the ERG i.e. $N$)
  \item Create $C$ disconnected cliques (clusters) on the diagonal, each one with size $k\times k$ where $k = n/C$ (we're creating a $GT$ made of balanced cliques).
\end{itemize}
This procedure outputs an unweighted adjacency matrix $G$ which represent an undirected unweighted Synthetic Graph.

\paragraph{Extension} Now we can extend this generation procedure to produce a wide spectrum of graphs, in particular let's consider these possibilities:
\begin{itemize}
\item Adding weights both to noise and to the structure.
\item Imbalance the clusters by varying the size of each one.
\item Introducing noise inside the clusters (we will refer to this noise as intra-noise or corrosion).
\end{itemize}

In this work not all the combination of the extensions have been analyzed since the spectrum is almost infinite, but the generation tool provides such possibility.

\subsection*{Experimental Setup}
In this section we will describe the setup of the experiments conducted and we will introduce the naming conventions used to report in the results.

\begin{figure}
\centering
\includegraphics[width=1\textwidth]{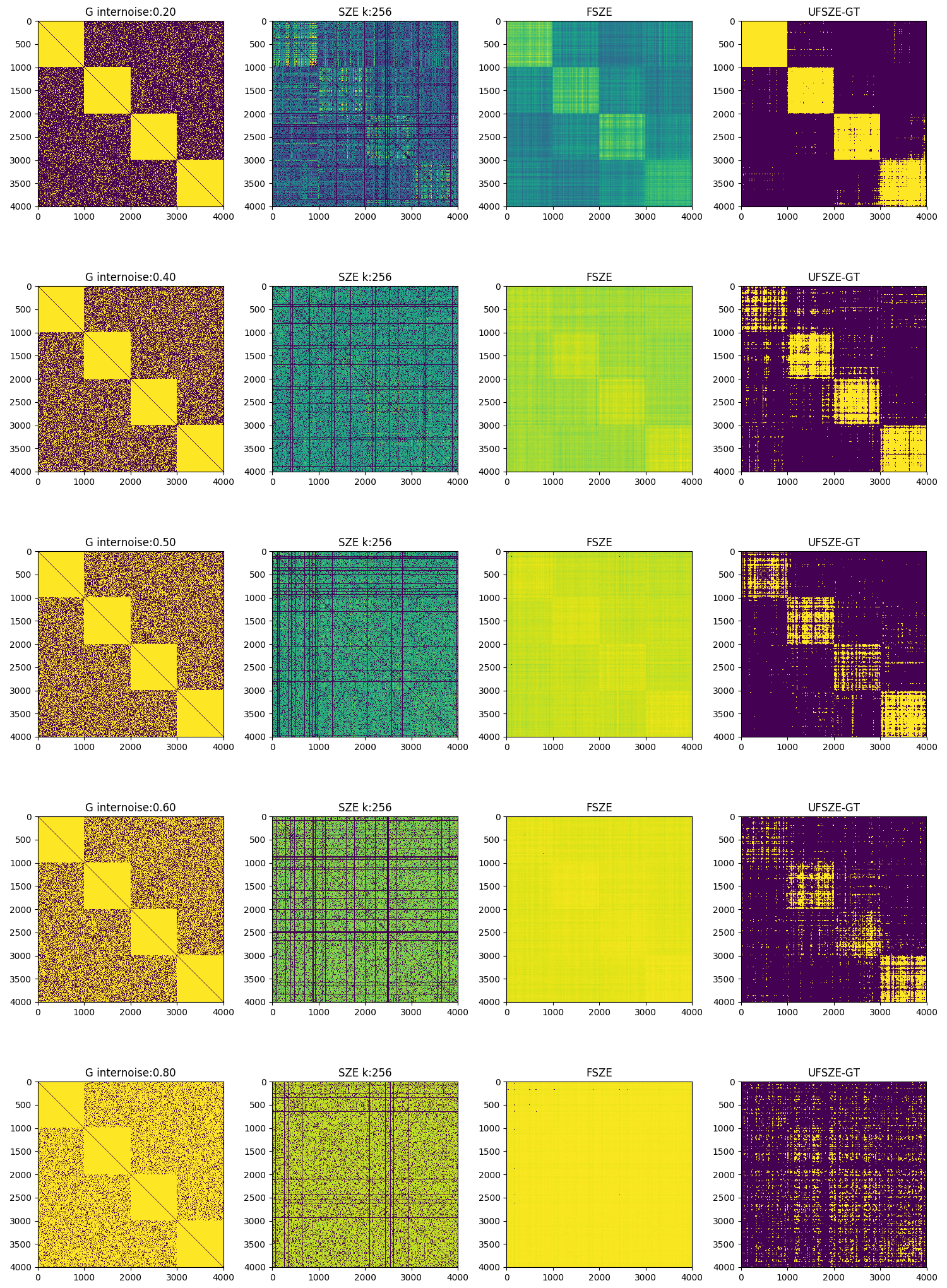}
\caption{\label{fig:20r}Visualization of a single repetition of Experiment 1}
\end{figure}

\paragraph{Experiment 1}
  \begin{itemize}
    \item For the first experiment several dimensions of the graphs have been considered. In particular we tried graphs with size $n=1000,2000,5000,10000,25000$.
    \item In particular, fixing a size $n$, we then generate 5 different graphs with $\text{inter-noise}=0.2, 0.4, 0.5, 0.6, 0.8$.
    \item We run the CoDec Procedure and calculate the measures specified in Section \ref{sec:measures} to validate the results.
    \item We performed a thresholding to unweight the FSZE matrix by minimizing this quantity:

$$\min_{0<t<1} l_2(FSZE_t,GT)$$

Which translated means that we tried to threshold the weights of the $FSZE$ matrix by minimizing $l_2(FSZE,GT)$ measure. This operation gives birth to the $UFSZE$ (Unweighted Filtered SZEmeredi matrix) (fourth column of the visual representation in Figure \ref{fig:20r}). This step has been inserted in order to study some possible relation between the values of the matrix produced by the CoDec Procedure and the density of the original graph. 
    \item The experiment has been repeated a certain number of times $r$ to generate some statistics. With medium size graphs the repetitions are 20 while with big graphs 5 or 3.
    \item We then visualize a single run
\end{itemize}
  
\paragraph{Experiment 2}
  \begin{itemize}
    \item For the second experiment we considered a fixed dimension for the synthetic graphs $n=4000$.
    \item Given a fixed size $n$, we generated imbalanced and balanced cluster-like graphs, by varying both internoise and intranoise, exploring all the combinations of these values: $0.2, 0.4, 0.5, 0.6, 0.8$.
    \item We run the CoDec Procedure and calculate the Adjusted Random Index measure specified in Section \ref{sec:measures} to validate the preservation of the structural properties.
    \item The experiment has been conducted one time, therefore there is no repetition.
    \item We then visualize a single run of a single graph.
\end{itemize}

\begin{figure}
\centering
\includegraphics[width=1\textwidth]{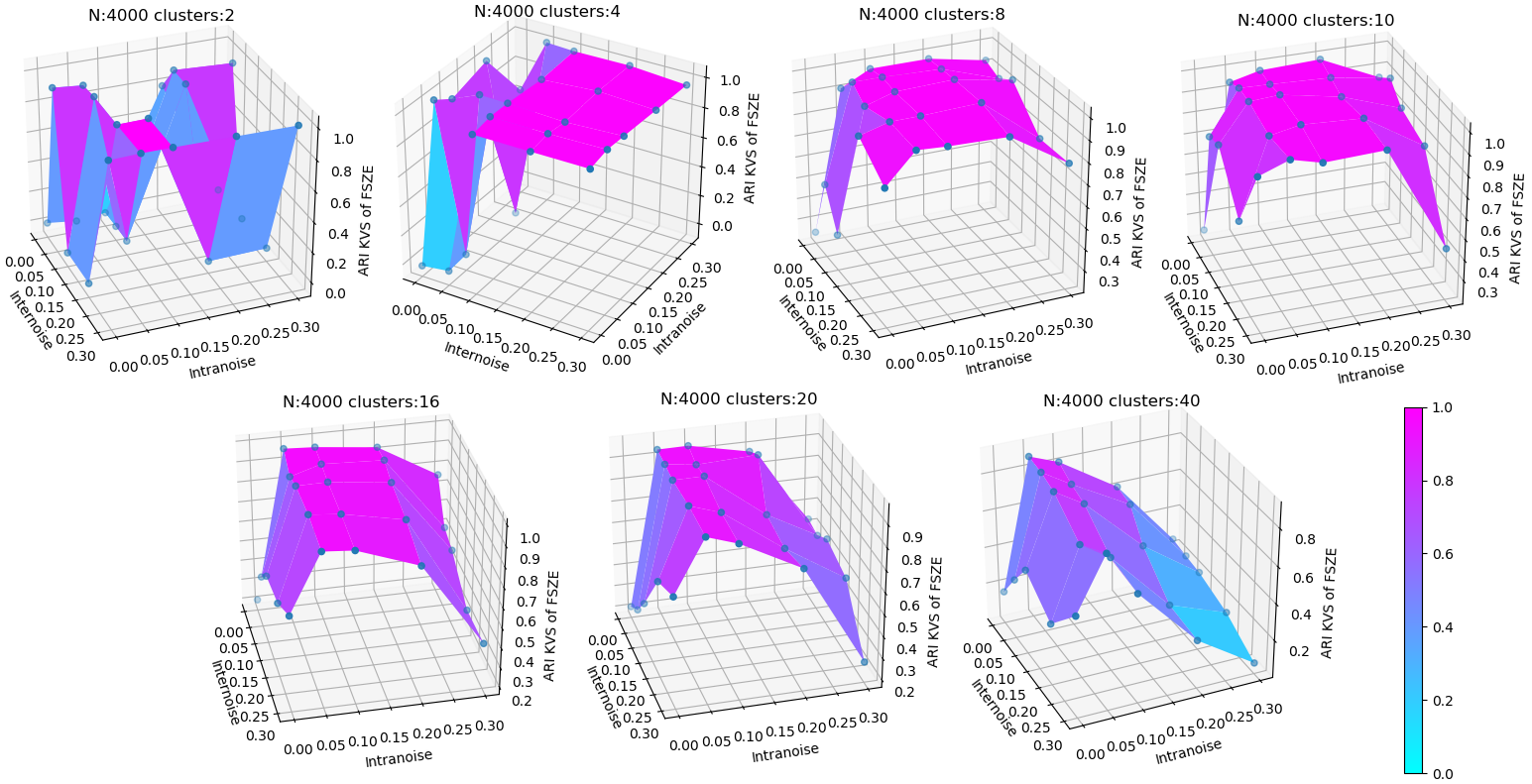}
\caption{\label{fig:imbalanced}Experiment 2: Internoise and Intranoise analysis over noisy graphs with imbalanced clusters as structure}
\end{figure}
   
\begin{figure}
\centering
\includegraphics[width=1\textwidth]{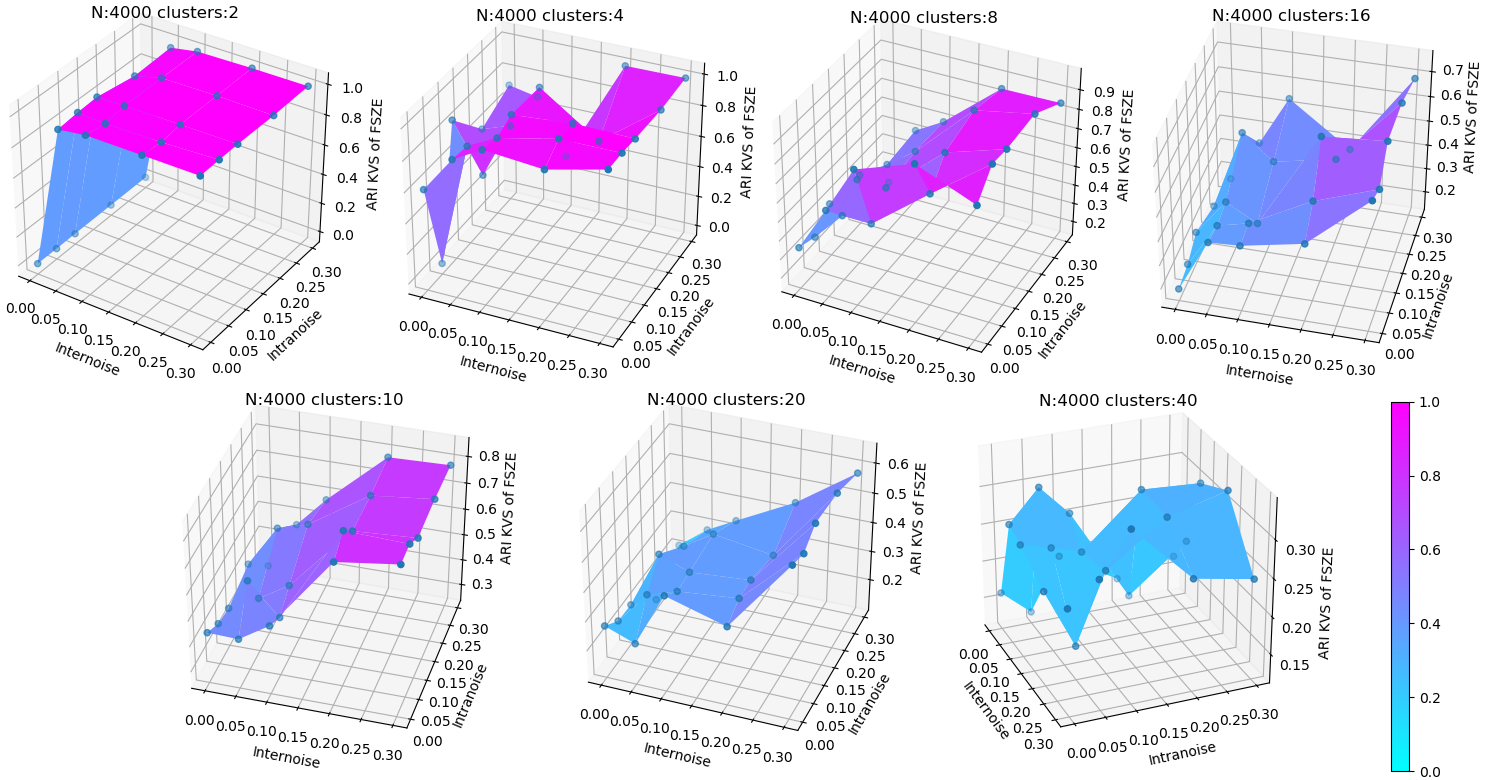}
\caption{\label{fig:balanced}Experiment 2: Internoise and Intranoise analysis over noisy graphs with balanced clusters as structure.}
\end{figure}

\input{tables/1000r20_tab}

\input{tables/2000r20_tab.tex}

\input{tables/5000r20_tab.tex}

\input{tables/10000r20_tab.tex}

\input{tables/25000r20_tab.tex}

\section{Real Network Experiments}\label{sec:realexperiments}
The majority of the experiments have been conducted with synthetic graphs because we have a ground truth to validate the structural preservation of the CoDec procedure. Since our framework is able to both approximate the given graph and highlight the underlying structural pattern we opted to test the first property with real world data. Although one of the requirements of the algorithm is to have a dense graph, we tested the devised procedure with real world datasets which, instead, are very sparse. The problem is that we don't have the ground truth and we can only calculate direct dissimilarity measures between the output matrix and the original one. The datasets have been taken from two famous repositories of real networks: the Stanford Large Network Dataset Collection SNAP \cite{snapnets} and from the Konect repository of the University Koblenz-Landau Konect \cite{konect}.

The setup of the experiments is very simple and we describe it as follows:
\begin{itemize}
\item All the networks used are unweighted and undirected.
\item We read the network in a matrix data structure and we apply the CoDec Procedure.
\item Since we don't have the ground truth to test the structural preservation we calculate the $l_2(G,FSZE)$ and the $l_1(G,FSZE)$ metrics to measure the dissimilarity of the two matrices.
\item Every network has been processed with 20 repetitions (just to show that the results does not significantly vary).
\end{itemize}

The table reports some statistics for each dataset, here is the naming convention:
    	\begin{itemize}
        \item k : the cardinality of the best partition, namely $|\mathcal{P}^*|$.
        \item $\epsilon$: the corresponding epsilon of the best partition found. 
        \item nirr: number of irregular pair of the final partition.
        \item $sze\_idx$: the Szemerédi Index described in Section \ref{sec:alon}.
        \item tcompress: time to compute the best partition and express it though $RED$ (step 1 and 2 of the CoDec Procedure described in Section \ref{sec:codec}). It is expressed in seconds.
        \item tdecompress: time to reconstruct the matrix and decompress the $RED$ matrix to the $SZE$ (step 3 of the CoDec Procedure described in Section \ref{sec:codec}). It is expressed in seconds.
        \item tfiltering: time to apply the median filter on the $SZE$ matrix giving $FSZE$ (step 4 of the CoDec Procedure described in Section \ref{sec:codec}).
    	\end{itemize}

\input{tables/movielens_table.tex}
\begin{figure}
\centering
\includegraphics[width=1\textwidth]{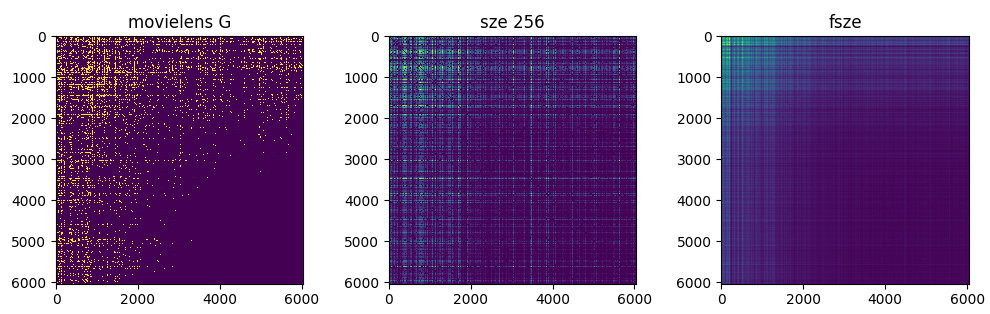}
\caption{\label{fig:movielens}Movielens \cite{movielens} network.}
\end{figure}

\input{tables/facebook_table.tex}
\begin{figure}
\centering
\includegraphics[width=1\textwidth]{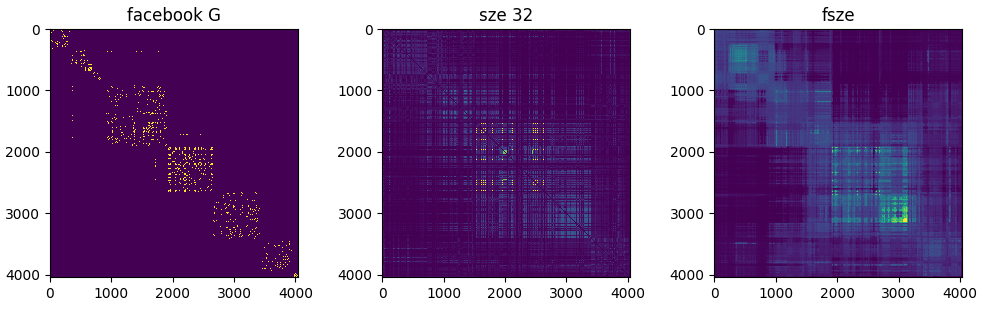}
\caption{\label{fig:facebook}Facebook \cite{facebookNIPS} network.}
\end{figure}

\input{tables/email_table.tex}
\begin{figure}
\centering
\includegraphics[width=1\textwidth]{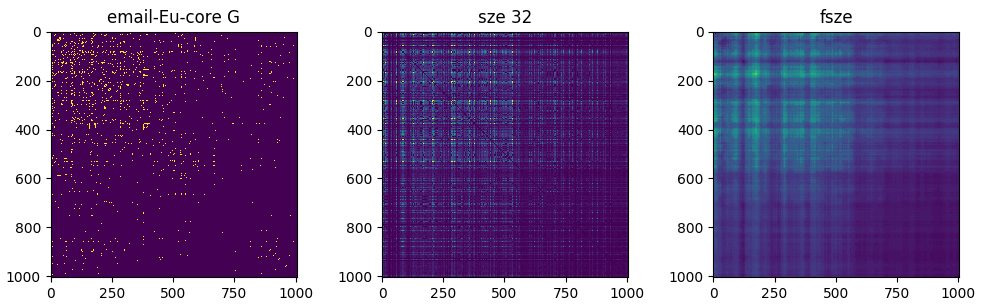}
\caption{\label{fig:email}Email-Eu-core \cite{Yin2017LocalClustering}\cite{Leskovec07email} network.}
\end{figure}

\input{tables/reactome_table.tex}
\begin{figure}
\centering
\includegraphics[width=1\textwidth]{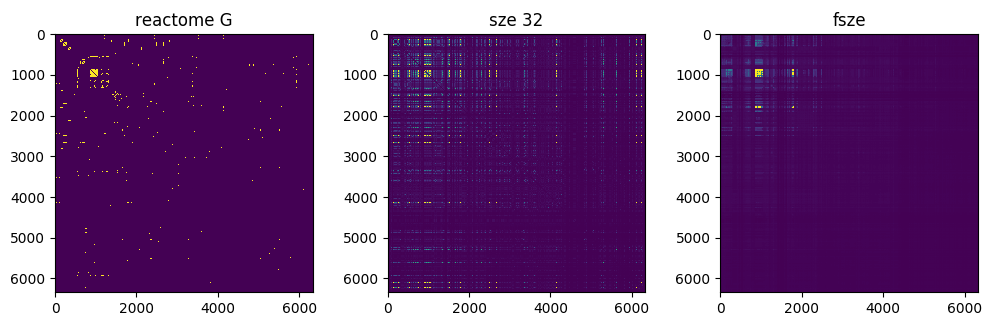}
\caption{\label{fig:reactome}Reactome \cite{Joshi-Tope2005Reactome:Pathways} network.}
\end{figure}

\section{Analysis and considerations} \label{sec:analysis}

\subsubsection*{Analysis of Synthetic Graphs Experiments}

The whole framework can be seen as a graph compression technique, beside the characteristic that allows us to store large graphs with less requirements, the procedure seems to preserve the underlying structural patterns which hold the information. 

As can be seen from all the tables, our compression algorithm based on an approximated version of Alon et al.'s algorithm plus the new devised heuristic, finds best partitions with epsilon value varying between $0.13$ to $0.27$. We recall that in order to have a faithful approximation of the original graph, we should be able to find partitions with very low values of epsilon. This is very difficult to achieve since one of the theoretical requirements is the astronomical size of the input graph (due to the tower constant on the cardinality of the partition). Of course even with a framework able to handle very big graphs we would still need a perfect implementation of Alon et al.'s algorithm. Said that, and keeping in mind that the epsilon parameter controls the approximation factor, we could say that it is a  good result. In fact we are able to process medium/big size graphs and achieve a very good compression with preservation of the structure.

As far as regards the compression property if we take in consideration Table \ref{tab:25000r20}, we can calculate the compression rate with graphs of 25000 nodes. In particular a $25000 \times 25000$ matrix has been compressed in a $2048 \times 2048$ matrix plus a $25000$ vector. If we weight each entry of a matrix and each entry of a vector as one byte (for simplicity) we have:
\begin{itemize}
\item Storage requirements for the graph: $(25000 \cdot 25000)/(2 \cdot 1024^{2}) \approx 298$ MB
\item Storage requirements for the compressed graph: $(2048 \cdot 2048)/(2 \cdot 1024^{2}) = 2$ MB
\item Storage requirements for the membership vector: $25000/1024 = 3.05$ KB
\end{itemize}

This constitutes a compression ratio of $147.25$ and we save $99.32\%$ of the space required. This is surely a very good result especially if we want to send the structure through a channel. This should suspect us since a compression rate such high it's uncommon, but this is a lossy compression and when the structure is decompressed it's meant to loose all the information which do not carry structural patterns, namely we only preserve useful informations.

From the visual inspection reported in Figure \ref{fig:20r} the CoDec Procedure first property is that it preserves the underlying structure of the graph $G$. This property should be stronger as the graph dimension grows, this point is also verified from the experimental results conducted with Experiment 1, in particular from the Adjusted Random Index of the KVS-clustering reported in Tables \ref{tab:1000r20}, \ref{tab:2000r20}, \ref{tab:5000r20}, \ref{tab:10000r20}, \ref{tab:25000r20}. We can clearly see that we achieve better results as the dimension of the graph increases, in fact with $n=25000$ we are still able to filter an inter-noise of $0.6$ with an Adjusted Random Index of $0.94$, while with $n=1000$ it is not the case. This confirms that as the dimension of the graph grows we are in fact capturing and preserving important information, therefore we can derive the fact that our approximated implementation of Alon et al.'s algorithm plus the new heuristic is indeed behaving according to the original formulation. 

Anther point is that Alon et al.'s algorithm does not tell us how to reach a regular partition at each iteration of the algorithm. In other words it does not fully specify the refinement step. Our new refinement heuristic has an underlying motivation which follows the combinatorial nature of the lemma and aims to mitigate the effect of the tower constant. In fact we find partitions with relatively small cardinality in medium/big graphs. The good side of this, is that we are able to compress the structure to very high rates and we are able to process small network too. There is also a bad side: since we have a low cardinality of the best partition it means that we will loose information in the decompression phase. That is because the internal structure of the partition classes is unknown and will be lost in the decompression phase. As we discussed in Section \ref{sec:decompression} we could keep track of the internal densities and later expand them through a random graph with probability of connection of two nodes equal to the internal density, but this has not been extensively tested and worth be further explored.

\begin{figure}
\centering
\includegraphics[width=1\textwidth]{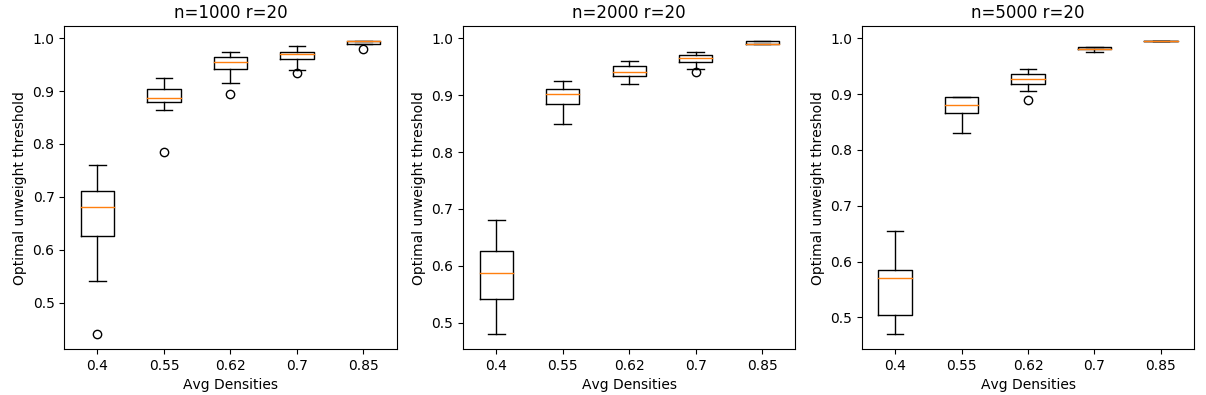}
\caption{\label{fig:empthreshold} Relation between graph density and optimal unweighting threshold, the graphs considered were the graphs used for the Experiment 1.}
\end{figure}

We then tried to study a possible relation between the density of the input graph and the optimal thresholding value of the $FSZE$ matrix which minimize the dissimilarity with the $GT$. The study have been reported in Figure \ref{fig:empthreshold}. There seems to be a pattern for high densities, but as it decrease the distribution of the optimal threshold value it's much wider. Maybe with more data and further investigations we can draw an empirical rule to optimally threshold the output of the CoDec procedure to produce an unweighted graph and to completely filter the noise by preserving only the structural part.\\

In Experiment 2 Figure \ref{fig:balanced} we tested the CoDec procedure robustness when both intranoise and internoise are added to a graph with balanced clusters as structure. We decided to use every combination of the values of the noise ranging from $0$ to $0.3$ since it represents the range where we achieve the best results and because of the computational time required. As said, we introduced both noisy edges between the clusters and corruption of the structure of the clusters. We can see that as the number of clusters increase the preservation is more difficult. This result can be explained by the fact that as the number of cliques increases the density of the original graph decreases, therefore we are getting far from the ideal conditions for the lemma to apply. Another important consideration is that the procedure is not able to preserve structure in total absence of noise, this is interesting since highlights the fact that the framework needs it to work and reinforce the denoisation property of the method. 

The same experiment has been conducted with imbalanced clusters, as reported in Figure \ref{fig:imbalanced}, one can immediately see that we have better results with respect to the balanced cluster case. The procedure preserves structure in many more combinations of intra and inter noise. We can see that as the intranoise increases (corruption of structure) we achieve worse preservation that is, again, due to the fact that as we corrupt we are in fact decreasing the density of the graph. It is not the case in the balanced cluster case, maybe because the critical threshold is higher due to the different morphology of the graphs.

\subsubsection*{Analysis of Real Networks Experiments}

Even if we assumed that the procedure needs a structural part of the graph make made of cluster-like groups, the algorithm is still able to highlight some of the pattern of the graphs. We can not provide a measure of preservation since we don't have a ground truth, but it can be verified from the visual inspection of the matrices in Figures \ref{fig:movielens}, \ref{fig:facebook}, \ref{fig:email}, \ref{fig:reactome}. This point is fundamental since it highlights the fact that this procedure is able to preserve the patterns of the graphs. In fact we could see it as if the graph $G$ is just a corroded version of a $GT$ which is well represented by the output of the procedure. 

Another important thing to notice is that even if we use graphs of medium size with a really low density, the framework is still able to find partitions and provide a visually appealing reconstruction with low $l_2$ dissimilarities. 

\begin{figure}
      \centering
      \includegraphics[width=0.4\textwidth]{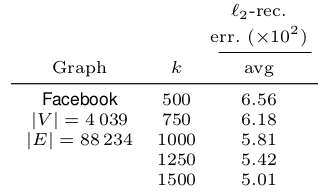}
      \caption{\label{fig:riondato}Riondato et al. $l_2$ reconstruction error for the Facebook network. We achieve 415.61.}
     \end{figure}

Perhaps the most important result obtained is the one that we got when the CoDec procedure has been applied to the Facebook network. In fact we obtain better results than the algorithm proposed by Riondato et al. \cite{Riondato2017} which, according to this survey \cite{Liu2017GraphSurvey}, is one of the algorithm at the sate-of-the-art. The result reported by Riondato at al. is a $l_2$ dissimilarity measure which is slightly different from our measure since it is not normalized by $n^2$ but instead by $n$. In order to compare the results we just need to multiply our dissimilarity value by the size of the graph, namely $n=4039$.
The results of Riondato et al. have been reported in Figure \ref{fig:riondato} in particular we achieve an $l_2=0.103 \cdot 4039=416.017$ while their best average is $501$ computed with 5 different runs. In order to fully compare the two techniques we should use the same graphs, unfortunately Riondato et al. tested their method with very big graphs that our developed framework can not yet handle. A workaround to test the two procedures would be to perform a sampling of the bigger graphs, but it requires an extensive study to select the best methodology to preserve the structural properties of the sampled graphs. 

Another important fact is that the algorithms of Riondato et al. mainly use graphs with clusters, and they point out that there is a relation between a weakly-regular partitions and the minimization of the CutNorm Error measure that they defined. In particular each class of the partition represent a cluster in the original graph. Our technique abstract this notion and in fact it is able also to highlight different patterns as we already discussed. 

Said that, we could strongly argue if the $l_2$ measure is a good proxy to measure how much structure has been preserved. In fact it is easy to see that as the graphs get more and more sparse we are closer to an empty graph which, unless our reconstruction adds a lot of noise, always gives low $l_2$ values. In Section \ref{sec:futurework} we discuss some other measures.




%% file: tables/1000r20_tab.tex
\begin{table}[t]
\centering
\begin{tabular}{cccc}
\hline
\multicolumn{4}{|c|}{\textbf{$n=1000$ $r=20$}} \\ \hline
\multicolumn{1}{|c|}{\textbf{Inter-noise}} & \textbf{Avg $\epsilon$} & \textbf{Avg k} & \multicolumn{1}{c|}{\textbf{Avg sze\_idx}} \\ \hline
\multicolumn{1}{|c|}{\textbf{0.2}} & 0.275 & 64 & \multicolumn{1}{c|}{0.0836} \\
\multicolumn{1}{|c|}{\textbf{0.4}} & 0.275 & 64 & \multicolumn{1}{c|}{0.1498} \\
\multicolumn{1}{|c|}{\textbf{0.5}} & 0.275 & 64 & \multicolumn{1}{c|}{0.1921} \\
\multicolumn{1}{|c|}{\textbf{0.6}} & 0.275 & 64 & \multicolumn{1}{c|}{0.2418} \\
\multicolumn{1}{|c|}{\textbf{0.8}} & 0.275 & 64 & \multicolumn{1}{c|}{0.3562} \\ \hline
\textbf{} &  &  &  \\ \hline
\multicolumn{1}{|c|}{\textbf{Inter-noise}} & \textbf{Avg ARI Kvs FSZE $\pm$ std} & \textbf{$l_2(FSZE,GT) \pm$ std} & \multicolumn{1}{c|}{\textbf{$l_1(FSZE,GT) \pm$ std}} \\ \hline
\multicolumn{1}{|c|}{\textbf{0.2}} & 0.9069 $\pm$ 0.01591 & 0.4119 $\pm$ 0.0007 & \multicolumn{1}{c|}{0.3928 $\pm$ 0.0004} \\
\multicolumn{1}{|c|}{\textbf{0.4}} & 0.9466 $\pm$ 0.00451 & 0.5016 & \multicolumn{1}{c|}{0.5009 $\pm$ 0.0009} \\
\multicolumn{1}{|c|}{\textbf{0.5}} & 0.8068 & 0.5543 & \multicolumn{1}{c|}{0.5467} \\
\multicolumn{1}{|c|}{\textbf{0.6}} & 0.7587 & 0.6071 & \multicolumn{1}{c|}{0.5862} \\
\multicolumn{1}{|c|}{\textbf{0.8}} & 0.0669 & 0.7299 & \multicolumn{1}{c|}{0.669} \\ \hline
\end{tabular}
\caption{Experiment 1: graphs with $n=1000$ and 20 repetitions. The standard deviation has been omitted in the cases where its below $10^{-5}$}
\label{tab:1000r20}

\end{table}

%% file: tables/2000r20_tab.tex
\begin{table}[t]
\centering

\begin{tabular}{cccc}
\hline
\multicolumn{4}{|c|}{\textbf{$n=2000$ $r=20$}} \\ \hline
\multicolumn{1}{|c|}{\textbf{Inter-noise}} & \textbf{Avg $\epsilon$} & \textbf{Avg k} & \multicolumn{1}{c|}{\textbf{Avg sze\_idx}} \\ \hline
\multicolumn{1}{|c|}{\textbf{0.2}} & 0.2975 & 128 & \multicolumn{1}{c|}{0.0912} \\
\multicolumn{1}{|c|}{\textbf{0.4}} & 0.275 & 128 & \multicolumn{1}{c|}{0.1515} \\
\multicolumn{1}{|c|}{\textbf{0.5}} & 0.275 & 128 & \multicolumn{1}{c|}{0.1942} \\
\multicolumn{1}{|c|}{\textbf{0.6}} & 0.275 & 128 & \multicolumn{1}{c|}{0.2434} \\
\multicolumn{1}{|c|}{\textbf{0.8}} & 0.275 & 128 & \multicolumn{1}{c|}{0.3586} \\ \hline
\textbf{} &  &  &  \\ \hline
\multicolumn{1}{|c|}{\textbf{Inter-noise}} & \textbf{Avg ARI Kvs FSZE $\pm$ std} & \textbf{$l_2(FSZE,GT) \pm$ std} & \multicolumn{1}{c|}{\textbf{$l_1(FSZE,GT) \pm$ std}} \\ \hline
\multicolumn{1}{|c|}{\textbf{0.2}} & 0.997 $\pm$ 0.0012 & 0.3764 $\pm$ 0.0006 & \multicolumn{1}{c|}{0.3549 $\pm$ 0.0006} \\
\multicolumn{1}{|c|}{\textbf{0.4}} & 0.9671 & 0.4972 & \multicolumn{1}{c|}{0.4966} \\
\multicolumn{1}{|c|}{\textbf{0.5}} & 0.8228 & 0.5543 & \multicolumn{1}{c|}{0.5468} \\
\multicolumn{1}{|c|}{\textbf{0.6}} & 0.7947 & 0.6086 & \multicolumn{1}{c|}{0.5876} \\
\multicolumn{1}{|c|}{\textbf{0.8}} & 0.0842 & 0.7299 & \multicolumn{1}{c|}{0.6686} \\ \hline
\end{tabular}
\caption{Experiment 1: graphs with $n=2000$ and 20 repetitions}
\label{tab:2000r20}
\end{table}

%% file: tables/5000r20_tab.tex
\begin{table}[t]
\centering
\begin{tabular}{cccc}
\hline
\multicolumn{4}{|c|}{\textbf{$n=5000$ $r=5$}} \\ \hline
\multicolumn{1}{|c|}{\textbf{Inter-noise}} & \textbf{Avg $\epsilon$} & \textbf{Avg k} & \multicolumn{1}{c|}{\textbf{Avg sze\_idx}} \\ \hline
\multicolumn{1}{|c|}{\textbf{0.2}} & 0.2435 & 256 & \multicolumn{1}{c|}{0.0936} \\
\multicolumn{1}{|c|}{\textbf{0.4}} & 0.2525 & 256 & \multicolumn{1}{c|}{0.1529} \\
\multicolumn{1}{|c|}{\textbf{0.5}} & 0.2525 & 256 & \multicolumn{1}{c|}{0.1958} \\
\multicolumn{1}{|c|}{\textbf{0.6}} & 0.23 & 256 & \multicolumn{1}{c|}{0.2445} \\
\multicolumn{1}{|c|}{\textbf{0.8}} & 0.23 & 256 & \multicolumn{1}{c|}{0.3599} \\ \hline
\textbf{} &  &  &  \\ \hline
\multicolumn{1}{|c|}{\textbf{Inter-noise}} & \textbf{Avg ARI Kvs FSZE $\pm$ std} & \textbf{$l_2(FSZE,GT) \pm$ std} & \multicolumn{1}{c|}{\textbf{$l_1(FSZE,GT) \pm$ std}} \\ \hline
\multicolumn{1}{|c|}{\textbf{0.2}} & 0.9981 $\pm$ 0.0028 & 0.3695 $\pm$ 0.0025 & \multicolumn{1}{c|}{0.3525 $\pm$ 0.0045} \\
\multicolumn{1}{|c|}{\textbf{0.4}} & 0.9935 $\pm$ 0.0024 & 0.5027 $\pm$ 0.0021 & \multicolumn{1}{c|}{0.5016 $\pm$ 0.0021} \\
\multicolumn{1}{|c|}{\textbf{0.5}} & 0.9857 & 0.5559 & \multicolumn{1}{c|}{0.547} \\
\multicolumn{1}{|c|}{\textbf{0.6}} & 0.6684 & 0.6164 & \multicolumn{1}{c|}{0.5944} \\
\multicolumn{1}{|c|}{\textbf{0.8}} & 0.2057 & 0.7362 & \multicolumn{1}{c|}{0.6714} \\ \hline
\end{tabular}
\caption{Experiment 1: graphs with $n=5000$ and 5 repetitions}
\label{tab:5000r20}

\end{table}

%% file: tables/10000r20_tab.tex
\begin{table}[t]
\centering
\begin{tabular}{cccc}
\hline
\multicolumn{4}{|c|}{\textbf{$n=10000$ $r=5$}} \\ \hline
\multicolumn{1}{|c|}{\textbf{Inter-noise}} & \textbf{Avg $\epsilon$} & \textbf{Avg k} & \multicolumn{1}{c|}{\textbf{Avg sze\_idx}} \\ \hline
\multicolumn{1}{|c|}{\textbf{0.2}} & 0.275 & 512 & \multicolumn{1}{c|}{0.0948} \\
\multicolumn{1}{|c|}{\textbf{0.4}} & 0.2525 & 512 & \multicolumn{1}{c|}{0.1533} \\
\multicolumn{1}{|c|}{\textbf{0.5}} & 0.2525 & 512 & \multicolumn{1}{c|}{0.1962} \\
\multicolumn{1}{|c|}{\textbf{0.6}} & 0.2525 & 512 & \multicolumn{1}{c|}{0.2449} \\
\multicolumn{1}{|c|}{\textbf{0.8}} & 0.23 & 512 & \multicolumn{1}{c|}{0.3607} \\ \hline
\textbf{} &  &  &  \\ \hline
\multicolumn{1}{|c|}{\textbf{Inter-noise}} & \textbf{Avg ARI Kvs FSZE $\pm$ std} & \textbf{$l_2(FSZE,GT) \pm$ std} & \multicolumn{1}{c|}{\textbf{$l_1(FSZE,GT) \pm$ std}} \\ \hline
\multicolumn{1}{|c|}{\textbf{0.2}} & 0.9993 $\pm$ 0.0003 & 0.3575 $\pm$ 0.0038 & \multicolumn{1}{c|}{0.3369 $\pm$ 0.0035} \\
\multicolumn{1}{|c|}{\textbf{0.4}} & 0.9907 & 0.5007 & \multicolumn{1}{c|}{0.5019} \\
\multicolumn{1}{|c|}{\textbf{0.5}} & 0.9918 & 0.558 & \multicolumn{1}{c|}{0.5476} \\
\multicolumn{1}{|c|}{\textbf{0.6}} & 0.7772 & 0.6228 & \multicolumn{1}{c|}{0.5942} \\
\multicolumn{1}{|c|}{\textbf{0.8}} & 0.194 & 0.7357 & \multicolumn{1}{c|}{0.6714} \\ \hline
\end{tabular}
\caption{Experiment 1: graphs with $n=10000$ and 5 repetitions}
\label{tab:10000r20}
\end{table}

%% file: tables/25000r20_tab.tex
\begin{table}[t]
\centering
\begin{tabular}{cccc}
\hline
\multicolumn{4}{|c|}{\textbf{$n=25000$ $r=3$}} \\ \hline
\multicolumn{1}{|c|}{\textbf{Inter-noise}} & \textbf{Avg $\epsilon$} & \textbf{Avg k} & \multicolumn{1}{c|}{\textbf{Avg sze\_idx}} \\ \hline
\multicolumn{1}{|c|}{\textbf{0.2}} & 0.32 & 2048 & \multicolumn{1}{c|}{0.0933} \\
\multicolumn{1}{|c|}{\textbf{0.4}} & 0.2975 & 2048 & \multicolumn{1}{c|}{0.1555} \\
\multicolumn{1}{|c|}{\textbf{0.5}} & 0.275 & 2048 & \multicolumn{1}{c|}{0.197} \\
\multicolumn{1}{|c|}{\textbf{0.6}} & 0.275 & 2048 & \multicolumn{1}{c|}{0.2459} \\
\multicolumn{1}{|c|}{\textbf{0.8}} & 0.2975 & 2048 & \multicolumn{1}{c|}{0.3615} \\ \hline
\textbf{} &  &  &  \\ \hline
\multicolumn{1}{|c|}{\textbf{Inter-noise}} & \textbf{Avg ARI Kvs FSZE $\pm$ std} & \textbf{$l_2(FSZE,GT) \pm$ std} & \multicolumn{1}{c|}{\textbf{$l_1(FSZE,GT) \pm$ std}} \\ \hline
\multicolumn{1}{|c|}{\textbf{0.2}} & 0.998 $\pm$ 0.0007 & 0.3599 $\pm$ 0.009 & \multicolumn{1}{c|}{0.3517 $\pm$ 0.0062} \\
\multicolumn{1}{|c|}{\textbf{0.4}} & 0.9995 & 0.4795 & \multicolumn{1}{c|}{0.481} \\
\multicolumn{1}{|c|}{\textbf{0.5}} & 0.9885 & 0.5755 & \multicolumn{1}{c|}{0.55} \\
\multicolumn{1}{|c|}{\textbf{0.6}} & 0.9482 & 0.6166 & \multicolumn{1}{c|}{0.5928} \\
\multicolumn{1}{|c|}{\textbf{0.8}} & 0.6862 & 0.6758 & \multicolumn{1}{c|}{0.6665} \\ \hline
\end{tabular}
\caption{Experiment 1: graphs with $n=25900$ and 3 repetitions}
\label{tab:25000r20}
\end{table}

%% file: tables/movielens_table.tex
\begin{table}[t]
\centering

\label{tab:movielens}
\begin{tabular}{ccccc}
\textbf{} & \multicolumn{4}{c}{\textbf{Movielens $n=6041$ $d=0.0541$ $r=20$}} \\ \cline{2-5} 
\multicolumn{1}{c|}{} & \textbf{k} & \textbf{$\epsilon$} & \textbf{sze\_idx} & \multicolumn{1}{c|}{\textbf{nirr}} \\ \cline{2-5} 
\multicolumn{1}{c|}{\textbf{min}} & 128 & 0.18200 & 0.00350 & \multicolumn{1}{c|}{599} \\
\multicolumn{1}{c|}{\textbf{avg}} & 217.6 & 0.23003 & 0.00386 & \multicolumn{1}{c|}{2457.2} \\
\multicolumn{1}{c|}{\textbf{sd}} & 60.2 & 0.02775 & 0.00015 & \multicolumn{1}{c|}{1348.7} \\
\multicolumn{1}{c|}{\textbf{max}} & 256 & 0.25000 & 0.00410 & \multicolumn{1}{c|}{3839} \\ \cline{2-5} 
\multicolumn{1}{c|}{} & \textbf{$l_2(SZE,G)$} & \textbf{$l_2(FSZE,G)$} & \textbf{$l_1(SZE,G)$} & \multicolumn{1}{c|}{\textbf{$l_1(FSZE,G)$}} \\ \cline{2-5} 
\multicolumn{1}{c|}{\textbf{min}} & 0.218 & 0.223 & 0.082 & \multicolumn{1}{c|}{0.071} \\
\multicolumn{1}{c|}{\textbf{avg}} & 0.219 & 0.224 & 0.084 & \multicolumn{1}{c|}{0.074} \\
\multicolumn{1}{c|}{\textbf{sd}} & 0.001 & 0.001 & 0.003 & \multicolumn{1}{c|}{0.003} \\
\multicolumn{1}{c|}{\textbf{max}} & 0.220 & 0.225 & 0.088 & \multicolumn{1}{c|}{0.078} \\ \cline{2-5} 
\multicolumn{1}{c|}{} & \textbf{tcompress/s} & \textbf{tdecompress/s} & \multicolumn{1}{c|}{\textbf{tfiltering/s}} & \multicolumn{1}{l}{} \\ \cline{2-4}
\multicolumn{1}{c|}{\textbf{min}} & 252.33 & 253.61 & \multicolumn{1}{c|}{473.55} & \multicolumn{1}{l}{} \\
\multicolumn{1}{c|}{\textbf{avg}} & 284.52 & 286.91 & \multicolumn{1}{c|}{511.04} & \multicolumn{1}{l}{} \\
\multicolumn{1}{c|}{\textbf{sd}} & 23.13 & 23.37 & \multicolumn{1}{c|}{29.67} & \multicolumn{1}{l}{} \\
\multicolumn{1}{c|}{\textbf{max}} & 321.99 & 325.05 & \multicolumn{1}{c|}{556.79} & \multicolumn{1}{l}{} \\ \cline{2-4}
\end{tabular}
\caption{Movielens real network}
\end{table}

%% file: tables/facebook_table.tex
\begin{table}[t]
\centering

\label{tab:facebook}
\begin{tabular}{ccccc}
\textbf{} & \multicolumn{4}{c}{\textbf{Facebook $n=4039$ $d=0.0108$ $r=20$}} \\ \cline{2-5} 
\multicolumn{1}{c|}{} & \textbf{k} & \textbf{$\epsilon$} & \textbf{sze\_idx} & \multicolumn{1}{c|}{\textbf{nirr}} \\ \cline{2-5} 
\multicolumn{1}{c|}{\textbf{min}} & 32 & 0.123 & 0.00010 & \multicolumn{1}{c|}{12} \\
\multicolumn{1}{c|}{\textbf{avg}} & 34.90 & 0.145 & 0.00015 & \multicolumn{1}{c|}{34.09} \\
\multicolumn{1}{c|}{\textbf{sd}} & 9.41 & 0.017 & 0.00007 & \multicolumn{1}{c|}{20.52} \\
\multicolumn{1}{c|}{\textbf{max}} & 64 & 0.191 & 0.00030 & \multicolumn{1}{c|}{100} \\ \cline{2-5} 
\multicolumn{1}{c|}{} & \textbf{$l_2(SZE,G)$} & \textbf{$l_2(FSZE,G)$} & \textbf{$l_1(SZE,G)$} & \multicolumn{1}{c|}{\textbf{$l_1(FSZE,G)$}} \\ \cline{2-5} 
\multicolumn{1}{c|}{\textbf{min}} & 0.101 & 0.103 & 0.018 & \multicolumn{1}{c|}{0.015} \\
\multicolumn{1}{c|}{\textbf{avg}} & 0.102 & 0.103 & 0.019 & \multicolumn{1}{c|}{0.016} \\
\multicolumn{1}{c|}{\textbf{sd}} & 0.001 & 0.000 & 0.000 & \multicolumn{1}{c|}{0.001} \\
\multicolumn{1}{c|}{\textbf{max}} & 0.103 & 0.103 & 0.020 & \multicolumn{1}{c|}{0.017} \\ \cline{2-5} 
\multicolumn{1}{c|}{} & \textbf{tcompress/s} & \textbf{tdecompress/s} & \multicolumn{1}{c|}{\textbf{tfiltering/s}} & \multicolumn{1}{l}{} \\ \cline{2-4}
\multicolumn{1}{c|}{\textbf{min}} & 43.07 & 43.27 & \multicolumn{1}{c|}{128.22} & \multicolumn{1}{l}{} \\
\multicolumn{1}{c|}{\textbf{avg}} & 46.55 & 46.78 & \multicolumn{1}{c|}{135.12} & \multicolumn{1}{l}{} \\
\multicolumn{1}{c|}{\textbf{sd}} & 2.23 & 2.24 & \multicolumn{1}{c|}{4.65} & \multicolumn{1}{l}{} \\
\multicolumn{1}{c|}{\textbf{max}} & 50.35 & 50.67 & \multicolumn{1}{c|}{145.48} & \multicolumn{1}{l}{} \\ \cline{2-4}
\end{tabular}
\caption{Facebook real network}
\end{table}

%% file: tables/email_table.tex
\begin{table}[t]
\centering

\label{tab:email}
\begin{tabular}{ccccc}
\textbf{} & \multicolumn{4}{c}{\textbf{Email-Eu-core $n=1005$ $d=0.00257$ $r=20$}} \\ \cline{2-5} 
\multicolumn{1}{c|}{} & \textbf{k} & \textbf{$\epsilon$} & \textbf{sze\_idx} & \multicolumn{1}{c|}{\textbf{nirr}} \\ \cline{2-5} 
\multicolumn{1}{c|}{\textbf{min}} & 16 & 0.131 & 0.00040 & \multicolumn{1}{c|}{3} \\
\multicolumn{1}{c|}{\textbf{avg}} & 30.4 & 0.219 & 0.00076 & \multicolumn{1}{c|}{29.1} \\
\multicolumn{1}{c|}{\textbf{sd}} & 4.9 & 0.027 & 0.00014 & \multicolumn{1}{c|}{10.7} \\
\multicolumn{1}{c|}{\textbf{max}} & 32 & 0.250 & 0.00100 & \multicolumn{1}{c|}{42} \\ \cline{2-5} 
\multicolumn{1}{c|}{} & \textbf{$l_2(SZE,G)$} & \textbf{$l_2(FSZE,G)$} & \textbf{$l_1(SZE,G)$} & \multicolumn{1}{c|}{\textbf{$l_1(FSZE,G)$}} \\ \cline{2-5} 
\multicolumn{1}{c|}{\textbf{min}} & 0.156 & 0.158 & 0.040 & \multicolumn{1}{c|}{0.035} \\
\multicolumn{1}{c|}{\textbf{avg}} & 0.157 & 0.158 & 0.041 & \multicolumn{1}{c|}{0.036} \\
\multicolumn{1}{c|}{\textbf{sd}} & 0.001 & 0.000 & 0.001 & \multicolumn{1}{c|}{0.001} \\
\multicolumn{1}{c|}{\textbf{max}} & 0.159 & 0.158 & 0.043 & \multicolumn{1}{c|}{0.039} \\ \cline{2-5} 
\multicolumn{1}{c|}{} & \textbf{tcompress/s} & \textbf{tdecompress/s} & \multicolumn{1}{c|}{\textbf{tfiltering/s}} & \multicolumn{1}{l}{} \\ \cline{2-4}
\multicolumn{1}{c|}{\textbf{min}} & 2.47 & 2.49 & \multicolumn{1}{c|}{8.07} & \multicolumn{1}{l}{} \\
\multicolumn{1}{c|}{\textbf{avg}} & 2.78 & 2.82 & \multicolumn{1}{c|}{8.50} & \multicolumn{1}{l}{} \\
\multicolumn{1}{c|}{\textbf{sd}} & 0.21 & 0.21 & \multicolumn{1}{c|}{0.25} & \multicolumn{1}{l}{} \\
\multicolumn{1}{c|}{\textbf{max}} & 3.18 & 3.22 & \multicolumn{1}{c|}{8.92} & \multicolumn{1}{l}{} \\ \cline{2-4}
\end{tabular}
\caption{Email-Eu-core real network}
\end{table}

%% file: tables/reactome_table.tex
\begin{table}[t]
\centering

\label{tab:reactome}
\begin{tabular}{ccccc}
\textbf{} & \multicolumn{4}{c}{\textbf{Reactome $n=6328$ $d=0.00073$ $r=20$}} \\ \cline{2-5} 
\multicolumn{1}{c|}{} & \textbf{k} & \textbf{$\epsilon$} & \textbf{sze\_idx} & \multicolumn{1}{c|}{\textbf{nirr}} \\ \cline{2-5} 
\multicolumn{1}{c|}{\textbf{min}} & 16 & 0.097 & 0.00010 & \multicolumn{1}{c|}{4} \\
\multicolumn{1}{c|}{\textbf{avg}} & 28.8 & 0.123 & 0.00011 & \multicolumn{1}{c|}{17.05} \\
\multicolumn{1}{c|}{\textbf{sd}} & 6.6 & 0.015 & 0.00002 & \multicolumn{1}{c|}{7.6} \\
\multicolumn{1}{c|}{\textbf{max}} & 32 & 0.165 & 0.00020 & \multicolumn{1}{c|}{31} \\ \cline{2-5} 
\multicolumn{1}{c|}{} & \textbf{$l_2(SZE,G)$} & \textbf{$l_2(FSZE,G)$} & \textbf{$l_1(SZE,G)$} & \multicolumn{1}{c|}{\textbf{$l_1(FSZE,G)$}} \\ \cline{2-5} 
\multicolumn{1}{c|}{\textbf{min}} & 0.082 & 0.083 & 0.011 & \multicolumn{1}{c|}{0.009} \\
\multicolumn{1}{c|}{\textbf{avg}} & 0.084 & 0.084 & 0.012 & \multicolumn{1}{c|}{0.010} \\
\multicolumn{1}{c|}{\textbf{sd}} & 0.001 & 0.001 & 0.000 & \multicolumn{1}{c|}{0.000} \\
\multicolumn{1}{c|}{\textbf{max}} & 0.085 & 0.085 & 0.013 & \multicolumn{1}{c|}{0.011} \\ \cline{2-5} 
\multicolumn{1}{c|}{} & \textbf{tcompress/s} & \textbf{tdecompress/s} & \multicolumn{1}{c|}{\textbf{tfiltering/s}} & \multicolumn{1}{l}{} \\ \cline{2-4}
\multicolumn{1}{c|}{\textbf{min}} & 97.84 & 98.28 & \multicolumn{1}{c|}{285.84} & \multicolumn{1}{l}{} \\
\multicolumn{1}{c|}{\textbf{avg}} & 105.49 & 105.96 & \multicolumn{1}{c|}{314.12} & \multicolumn{1}{l}{} \\
\multicolumn{1}{c|}{\textbf{sd}} & 3.82 & 3.84 & \multicolumn{1}{c|}{12.87} & \multicolumn{1}{l}{} \\
\multicolumn{1}{c|}{\textbf{max}} & 110.89 & 111.38 & \multicolumn{1}{c|}{330.39} & \multicolumn{1}{l}{} \\ \cline{2-4}
\end{tabular}
\caption{Reactome real network}
\end{table}

%% file: conclusion.tex
In this thesis it has been devised a new pipeline for compressing and decompressing a graph by exploiting powerful Szemerédi's Regularity Lemma. In particular, it has been developed a procedure called CoDec (Compression-Decompression) that uses an approximated algorithmic formulation of the strong Regularity Lemma given by Alon et al. as compression step, then after the decompression phase (which simply exploits the Embedding Lemma) it introduces a post-decompression step to better highlight the structural properties of the reconstructed graph.

We provided a new heuristic for the refinement step which exploits the internal structures of the classes of a regular partition. As the experiments show, the new heuristic mitigates the problematic tower constant on the number of classes which should compose a regular partition. In fact when prompted with sufficiently dense graphs, the framework always find regular partitions of different cardinalities. 

We then provided an extensive experimental evaluation to measure how robust is the framework as we both corrupt the structures of the synthetic graphs, which carries the information, and add noisy edges between them. As we reported, the procedure is less robust to the corruption of the structure since it implies a decrement in the density of the graph. This is coherent to the requirements to the Regularity Lemma which finds its ideal usage in dense graphs. In fact, CoDec achieves good structural preservation properties when we add noise between the structures. 

We also tested the pipeline as the graphs dimension increase. Turns out that our implementation seems to confirm the theoretical requirements needed by the strong Regularity Lemma, so by processing bigger graphs we achieve better results.  

\section{Future Works}\label{sec:futurework}

In this Section we report some ideas that can be further explored in the future. In particular, we could study how the results of the procedure change when the threshold of the Sparsification and Densification branch of the new heuristic vary. This could be another experiment to better understand the properties of the new heuristic and could possibly open new insights.

Surely we should also perform several tests on weighted graphs. In particular, we could study if there is any relation between weighted internoise and intranoise of the input graph and the CoDec structural preservation ability. The introduction of weighted noise and the introduction of weighted structures could open new aspects of the algorithm both in positive and negative.

Another very important aspect would be to use different technologies to implement the framework. The whole pipeline has been coded using Python 3.6 (with heavy usage of the Numpy library). We have some limitations on the dimension of the graphs that we can process due to the choice of using matrices as data structure representation for the graphs. We could use different data structures and we could of course make use of some dedicated libraries like \href{https://graph-tool.skewed.de/}{GraphTools} or \href{https://networkx.github.io/}{NetworkX} which are simply born for such applications. Another option would be to port the framework into another language like C++ to achieve a speedup on the running time, but the migration would not be immediate due to the complexity of the project and target language. A big improvement would be to redefine the framework in a distributed environment, this surely open new possibility to process bigger graphs and to explore different refinements. 

As we said in the analysis of the result, the $l_2$ metric does not provide a good evaluation to compare the similarity of two graphs, this is always the case when we are evaluating techniques that offers an approximation of the graph. In particular, we would like a measure of how similar are the two structural part of the graphs considered. An idea would be to use an indirect measure based on a feature vector of some statistics of the graphs, like: average local clustering coefficient from Watts et al. \cite{Watts1998CollectiveNetworks}, global clustering coefficient from Luce et al. \cite{Luce1949AStructure}, average degree and so on, and then measure the distance between the two feature vectors. This is an indirect measure of similarity, the problem of computing such feature vector is that we need more computational time and it may be a problem when dealing with huge graphs. 

There is also the possibility to use the spectral norm that is defined as the square root of the maximum eigenvalue of the squared adjacency matrix. Since it combines the squared adjacency matrix (which counts the number of paths of length two), and eigenvalues (which are highly connected with spectral clustering), it could be a good proxy to evaluate the structural similarity of two (cluster like) graphs.  

In this work we used the notion of strong regularity to compress an input graph while preserving some structural properties. However, we can relax the notion to produce weak regular partition exploiting Frieze and Kannan's \cite{Frieze1999APartition} weak version of the Regularity Lemma. There would be some benefits to move from a strong to a weak regularity notion mainly because we are no longer required to work with very big graphs and because there are many papers continuously improving the tower bounds and providing new approximation algorithms like the last by Fox et al. \cite{Fox2018ARegularity}.

%% file: main.bbl
\begin{thebibliography}{10}

\bibitem{Liu2017GraphSurvey}
Y.~Liu, A.~Dighe, T.~Safavi, and D.~Koutra, ``{Graph Summarization: A
  Survey},'' {\em ACM Computing Surveys}, vol.~V, no.~N, 2017.

\bibitem{SperottoPelillo}
A.~Sperotto and M.~Pelillo, ``{Szemer{\'{e}}di's Regularity Lemma and Its
  Applications to Pairwise Clustering and Segmentation},'' in {\em Energy
  Minimization Methods in Computer Vision and Pattern Recognition} (A.~L.
  Yuille, S.-C. Zhu, D.~Cremers, and Y.~Wang, eds.), (Berlin, Heidelberg),
  pp.~13--27, Springer Berlin Heidelberg, 2007.

\bibitem{Alon94}
N.~Alon, R.~A. Duke, H.~Lefmann, V.~R{\"{o}}dl, and R.~Yuster, ``{The
  Algorithmic Aspects of the Regularity Lemma},'' {\em J. Algorithms}, vol.~16,
  pp.~80--109, 1 1994.

\bibitem{PelilloRevealing17}
M.~Pelillo, I.~Elezi, and M.~Fiorucci, ``Revealing structure in large graphs:
  Szemerédi’s regularity lemma and its use in pattern recognition,'' {\em
  Pattern Recognition Letters}, vol.~87, pp.~4 -- 11, 2017.
\newblock Advances in Graph-based Pattern Recognition.

\bibitem{Fioruccietal}
M.~Fiorucci, A.~Torcinovich, M.~Curado, F.~Escolano, and M.~Pelillo, ``{On the
  Interplay Between Strong Regularity and Graph Densification},'' in {\em
  Graph-Based Representations in Pattern Recognition} (P.~Foggia, C.-L. Liu,
  and M.~Vento, eds.), (Cham), pp.~165--174, Springer International Publishing,
  2017.

\bibitem{Gowers1997}
W.~T. Gowers, ``{Lower bounds of tower type for Szemer{\'{e}}di's uniformity
  lemma},'' {\em Geometric {\&} Functional Analysis GAFA}, vol.~7,
  pp.~322--337, 5 1997.

\bibitem{Szemeredi1969OnProgression}
E.~Szemer{\'{e}}di, ``{On sets of integers containing no four elements in
  arithmetic progression},'' {\em Acta Mathematica Academiae Scientiarum
  Hungaricae}, vol.~20, no.~1-2, pp.~89--104, 1969.

\bibitem{Furstenberg1982TheTheorem}
H.~Furstenberg, Y.~Katznelson, and D.~Ornstein, ``{The Ergodic theoretical
  proof of Szemer{\'{e}}di’s theorem},'' {\em Bulletin of the American
  Mathematical Society}, vol.~7, no.~3, pp.~527--552, 1982.

\bibitem{Gowers1998}
W.~T. Gowers, ``{A New Proof of Szemer{\'{e}}di's Theorem for Arithmetic
  Progressions of Length Four},'' {\em Geometric {\&} Functional Analysis
  GAFA}, vol.~8, pp.~529--551, 7 1998.

\bibitem{Komlos1996SzemeredisTheory}
J.~Komlos and M.~Simonovits, ``{Szemer{\'{e}}di's Regularity Lemma and its
  applications in graph theory},'' {\em DIMACS Technical Report}, 1996.

\bibitem{Komlos2002TheTheory}
J.~Koml{\'{o}}s, A.~Shokoufandeh, M.~Simonovits, and E.~Szemer{\'{e}}di, ``{The
  Regularity Lemma and Its Applications in Graph Theory},'' {\em Theoretical
  aspects of computer science (Tehran, 2000)}, vol.~2292, pp.~84--112, 2002.

\bibitem{Szemeredi75Graphs}
E.~Szemeredi, ``{Regular Partitions of Graphs.},'' tech. rep., Stanford, CA,
  USA, 1975.

\bibitem{Frieze1999APartition}
A.~Frieze and R.~Kannan, ``{A simple algorithm for constructing
  Szemer{\'{e}}di's Regularity Partition},'' 1999.

\bibitem{Neil2014ExperimentalClustering}
K.~O. Neil and S.~L. Peters, ``{Experimental Improvements to Regularity
  Clustering},'' tech. rep., 2014.

\bibitem{Szeliski2010ComputerApplications}
R.~Szeliski, ``{Computer Vision : Algorithms and Applications},'' {\em
  Computer}, vol.~5, p.~832, 2010.

\bibitem{Hubert1985}
L.~Hubert and P.~Arabie, ``{Comparing partitions},'' {\em Journal of
  Classification}, vol.~2, pp.~193--218, 12 1985.

\bibitem{snapnets}
J.~Leskovec and A.~Krevl, ``{SNAP Datasets}: {Stanford} large network dataset
  collection.'' \url{http://snap.stanford.edu/data}, June 2014.

\bibitem{konect}
J.~Kunegis, ``{KONECT} -- {The} {Koblenz} {Network} {Collection},'' in {\em
  Proc. Int. Conf. on World Wide Web Companion}, pp.~1343--1350, 2013.

\bibitem{movielens}
{GroupLens Research}, ``{MovieLens} data sets.''
  \url{http://www.grouplens.org/node/73}, October 2006.

\bibitem{facebookNIPS}
J.~Leskovec and J.~J. Mcauley, ``{Learning to Discover Social Circles in Ego
  Networks},'' in {\em Advances in Neural Information Processing Systems 25}
  (F.~Pereira, C.~J.~C. Burges, L.~Bottou, and K.~Q. Weinberger, eds.),
  pp.~539--547, Curran Associates, Inc., 2012.

\bibitem{Yin2017LocalClustering}
H.~Yin, A.~R. Benson, J.~Leskovec, and D.~F. Gleich, ``{Local Higher-Order
  Graph Clustering},'' in {\em Proceedings of the 23rd ACM SIGKDD International
  Conference on Knowledge Discovery and Data Mining - KDD '17}, pp.~555--564,
  2017.

\bibitem{Leskovec07email}
J.~Leskovec, J.~Kleinberg, and C.~Faloutsos, ``{Graph Evolution: Densification
  and Shrinking Diameters},'' {\em ACM Trans. Knowl. Discov. Data}, vol.~1, 3
  2007.

\bibitem{Joshi-Tope2005Reactome:Pathways}
G.~Joshi-Tope, M.~Gillespie, I.~Vastrik, P.~D{\&}apos;Eustachio, E.~Schmidt,
  B.~de~Bono, B.~Jassal, G.~R. Gopinath, G.~R. Wu, L.~Matthews, S.~Lewis,
  E.~Birney, and L.~Stein, ``{Reactome: A knowledgebase of biological
  pathways},'' {\em Nucleic Acids Research}, vol.~33, no.~DATABASE ISS., 2005.

\bibitem{Riondato2017}
M.~Riondato, D.~Garcia-Soriano, and F.~Bonchi, ``{Graph Summarization with
  Quality Guarantees},'' {\em Data Min. Knowl. Discov.}, vol.~31, pp.~314--349,
  3 2017.

\bibitem{Watts1998CollectiveNetworks}
D.~J. Watts and S.~H. Strogatz, ``{Collective dynamics of 'small-world'
  networks},'' {\em Nature}, vol.~393, no.~6684, pp.~440--442, 1998.

\bibitem{Luce1949AStructure}
R.~D. Luce and A.~D. Perry, ``{A method of matrix analysis of group
  structure},'' {\em Psychometrika}, vol.~14, no.~2, pp.~95--116, 1949.

\bibitem{Fox2018ARegularity}
J.~Fox, L.~M. Lov{\'{a}}sz, and Y.~Zhao, ``{A fast new algorithm for weak graph
  regularity},'' pp.~1--13, 2018.

\end{thebibliography}
